\documentclass[journal]{IEEEtran}

\usepackage{graphicx} 
\usepackage{float} 
\usepackage{cite} 
\usepackage{amsmath, amsfonts, amssymb}
\usepackage{algorithmic}
\usepackage{algorithm}
\usepackage{enumerate}
\usepackage{tabularx}
\usepackage{array}
\usepackage{comment}
\usepackage{xcolor,ulem} 
\usepackage[caption=false]{subfig}
\def\slu#1{\textcolor{black}{#1}}

\begin{document}
\title{Deep-Learning-Aided Successive Cancellation List Flip Decoding for Polar Codes}

\author{Fu-Siang~Liang, Shan~Lu,~\IEEEmembership{Member,~IEEE} and
		Yeong-Luh~Ueng,~\IEEEmembership{Senior Member,~IEEE}
		
\thanks{Fu-Siang~Liang was with the Dept. of Electrical Engineering, National Tsing Hua University, Hsinchu, Taiwan. (e-mail: 868503@gmail.com).}
\thanks{Shan~Lu is with the Department of Information and Communication Engineering, Graduate School of Engineering, Nagoya University, Nagoya, Japan. (e-mail: shan.lu.jp@ieee.org)}
\thanks{Yeong-Luh~Ueng is with the Dept. of Electrical Engineering and the Institute of Communications Engineering, National Tsing Hua University, Hsinchu, Taiwan. (e-mail: ylueng@ee.nthu.edu.tw).}
}

\markboth{IEEE Transactions on Cognitive Communications and Networking}{}

\maketitle
\begin{abstract}
Polar codes are the first error-correcting code proven to achieve channel capacity based on infinite code length.
The Successive Cancellation List Flip (SCLF) decoding algorithm was proposed by flipping an erroneous bit during the next decoding attempt.
To identify the erroneous bits, the Log-Likelihood Ratio (LLR) is used to indicate the reliability of each decision bit.
To improve the accuracy of the erroneous bit prediction, we propose deep-learning-aided (DL-aided) SCLF decoding algorithms.
We first offer a stacked LSTM network that contains new features to train our models, which are able to improve the accuracy of the prediction of positions of erroneous bits.
Then we separately train the stacked LSTM models to predict the position of both the first and second erroneous bits and whether to continue flipping.
As a result, the DL-aided SCLF decoding algorithms based on the proposed stacked LSTM \mbox{flip-1} model, stacked LSTM \mbox{flip-2} model, and \slu{the stacked LSTM \mbox{continue-flipping} check (CFC) model} are able to provide a better performance at a lower number of average decoding attempts when compared to other state-of-the-art decoding algorithms.

\par
\vspace{0.1in} \emph{Index Terms} --- Polar codes, successive cancellation list (SCL) decoding, long short-term memory (LSTM), deep learning
\end{abstract}

\section{Introduction}
\IEEEPARstart{A}{s} a class of error-correcting code, polar codes have been proposed in order to achieve the theoretical channel capacity by Arıkan in \cite{Arikan}.
Arıkan proposed the low-complexity successive cancellation (SC) algorithm for decoding.
Since the error-correcting performance of SC decoding is insufficient for finite-length polar codes, the successive cancellation list (SCL) decoding method was proposed in \cite{SCL_1, SCL_2}.
This method monitors up to $L$ decoding paths during the process.
Additionally, the Cyclic Redundancy Check (CRC) aided SCL (CA-SCL) decoding method \cite{CASCL} was introduced as an enhancement to the SCL decoding method, which uses a CRC to aid in the selection of the final output path. 
\slu{In order to reduce the decoding latency and achieve a higher throughput, Fast-SC decoding was presented in \cite{SSC} and \cite{FastPolar} that includes different special nodes, such as the \mbox{Rate-0} node, the \mbox{Rate-1} node, the \mbox{Single-Parity-Check} (SPC) node, and the repetition (REP) node.} 

The SC decoding algorithm, however, is prone to severe error propagation, where a single incorrect bit can affect the accuracy of subsequent decoded information bits in the decoded codeword. To mitigate this issue, the Successive Cancellation Flip (SCF) decoding algorithm \cite{SCF_basic} was proposed.
This method flips the channel-induced erroneous bit during the following decoding attempt.
Log-Likelihood Ratio (LLR) is used as a metric to indicate the reliability of each decision bit \cite{SCF_basic}.
Several works have further optimized the metric to achieve an improved performance \cite{SCF_single_bit_1, SCF_single_bit_2}.
Additionally, other methods \cite{SCF_multi_bit_1, SCF_multi_bit_2, SCF_multi_bit_3} have been developed that simultaneously flip multiple bits since channel-induced errors sometimes occur in more than one position.

\slu{
However, the perturbation parameter used in the flipping metric of \cite{SCF_multi_bit_3} introduces transcendental computations, leading to increased implementation complexity.
Consequently, \cite{DSCF_ml} proposed an alternative perturbation parameter that is added to the magnitude of LLR values in the flip metric formula to avoid transcendental computations.
In addition, the new perturbation parameter is further optimized by the machine learning technique, even with a limited training dataset, to achieve improved performance.
To preserve the benefits of the flipping scheme and low latency of Fast-SC decoding, \cite{FSCF_rl} proposed an algorithm that generates a specific vector of LLR values for Fast-SC decoding as a preliminary step before deriving the flipping metric.
Additionally, a perturbation parameter is proposed to be applied to the flipping metric, which is optimized using reinforcement learning techniques.
}

The Successive Cancellation List Flip (SCLF) decoding algorithm, first proposed in \cite{CASCLF_basic_1}, combines the concepts of SCF decoding with CA-SCL decoding.
In \cite{CASCLF_basic_2}, a new metric was proposed that identifies the positions of any channel-induced erroneous bits.
However, this method only allows for flipping a single bit during each decoding attempt.
Later, more reliable metrics were proposed in \cite{CASCLF_multi_bit_1, CASCLF_multi_bit_2} that support simultaneous flipping of multiple bits during each decoding attempt.
The methods presented in \cite{CASCLF_basic_2, CASCLF_multi_bit_1, CASCLF_multi_bit_2} are referred to as the \textit{metric-based sorting method} in this paper.

\slu{
The authors of \cite{CASCLF_ml_4} and \cite{CASCLF_ml_5} propose to flip erroneous bits based on the flipping metric rather than the machine learning model prediction.
In \cite{CASCLF_ml_4}, an improved flip metric is proposed to identify erroneous bits, which is derived from \cite{SCF_multi_bit_3}.
The proposed flipping metric exploits the inherent relations among the information bits in polar codes, represented by a correlation matrix to further improve the error correction performance.
Furthermore, the value of the correlation matrix is optimized based on the technique of stochastic gradient descent (SGD).
The fast successive-cancellation list flip (Fast-SCLF) decoding algorithm proposed in \cite{CASCLF_ml_5} is similar to the method in \cite{FSCF_rl}.
However, \cite{CASCLF_ml_5} optimizes a perturbation parameter using SGD technique, similar to \cite{CASCLF_ml_4}, instead of reinforcement learning in \cite{FSCF_rl}.
Another distinction between the two algorithms is that one is applied to SCL decoding, while the other is applied to SC decoding.
Both algorithms aim to reduce the decoding complexity by utilizing special nodes.
}

To locate the erroneous bit positions generated through SCLF decoding algorithms, several deep-learning-aided (DL-aided) SCLF decoding methods have been proposed based on the Long Short-Term Memory (LSTM) model, such as those described in \cite{SCF_ml_1, SCF_ml_2, CASCLF_ml_1, CASCLF_ml_2, CASCLF_ml_3}.
The LSTM model replaces the metric-based sorting method to achieve a higher probability of identifying the position of an erroneous bit.

The models in \cite{SCF_ml_1, SCF_ml_2, CASCLF_ml_2}, and \cite{CASCLF_ml_3} are trained using the LLR sequence from the decoding attempt that initially failed.
A simplified metric, as a feature for training the LSTM model to achieve a higher probability of identifying the position of an erroneous bit, was presented in \cite{CASCLF_ml_1}.
However, these methods are limited to flipping only a single bit in each decoding attempt.
The LSTM is used to predict the presence of an erroneous bit and decide whether to perform a two-bit flipping decoding attempt \cite{SCF_ml_1}.
However, the prediction of the second erroneous bit is based on the LLR sequence from the failure of the initial decoding.
Hence, the prediction for the second erroneous bit is inaccurate since the decoding trajectory is changed after the first erroneous bit is flipped.

Therefore, to improve the accuracy of the prediction of the erroneous bit, we consider a DL-aided SCLF decoding algorithm based on separately trained models to predict the positions of both the first and the second erroneous bit and the decision of whether to continue flipping.

In summary, the main contributions of this paper are:
\begin{enumerate}
    \item A stacked LSTM network with new features to train our models, which can improve prediction accuracy.
    \item Separately training the models to predict the positions of the first (stacked LSTM flip-1 model) and the second (stacked LSTM flip-2 model) erroneous bit and the decision of whether to continue flipping to improve the prediction accuracy (\slu{the stacked LSTM continue-flipping check (CFC) model}).
    \item The DL-aided SCLF decoding algorithms that include stacked LSTM networks are considered. The DL-SCLF-1 decoding method is based on basic SCLF decoding and uses a stacked LSTM flip-1 model to predict the position of the first erroneous bit. The DL-SCLF-2 decoding method is a combination of the stacked LSTM flip-1 model, the stacked LSTM flip-2 model, and the stacked LSTM CFC model.
    \item Simulation results show that the \slu{frame error rate} (FER) performance and complexity in terms of  average decoding attempts (ADA) of our proposed algorithms are improved in comparison with the existing SCLF decoding algorithms presented in \cite{CASCLF_basic_2, CASCLF_multi_bit_1, CASCLF_ml_1}.
\end{enumerate}

The remainder of the paper is organized as follows:
Section II discusses the preliminaries for polar codes and decoding algorithms.
Section III demonstrates the proposed neural networks for SCLF decoding.
Section IV presents the proposed SCLF decoding algorithms.
Section V shows the experimental results.
Finally, we conclude the work in Section VI.

\section{Preliminaries}

\subsection{Construction of Polar Codes}
An $(N, K)$ polar code constructed based on the channel polarization is a linear block code, where the code length is $N=2^n$, $n\in\mathbb{Z}^+$, and the number of information bits allocated in the reliable bit channels is $K$.
The code rate is $R=K/N$.
The remaining $N-K$ bits, called frozen bits, are predefined as zero and are allocated to the unreliable bit channels.
\slu{The reliability of each bit channel is determined by the polarization weight \cite{PW} in our work.}
The set of information bit indices is denoted by $\mathcal{A}$, and the set of frozen bit indices is denoted by $\mathcal{A}^c$.

The codeword vector is represented by $x^N_1 = (x_1, x_2, \ldots, x_N)$ and the message vector is represented by $u^N_1 = (u_1, u_2, \ldots, u_N)$. The polar code is encoded as
\begin{equation}
x_1^N= u_1^N\mathbf{G}_N,
\end{equation}
where $\mathbf{G}_N$ is the generator matrix defined as $\mathbf{G}_N=\mathbf{B}_N\mathbf{F}^{\otimes n}$.
Here $\mathbf{B}_N$ represents the bit-reversal permutation matrix, and $\mathbf{F}^{\otimes n}$ is the $n$-th Kronecker power of $\mathbf{F}$, where $\mathbf{F} \triangleq [
    \begin{smallmatrix}
    1 & 0 \\
    1 & 1
    \end{smallmatrix}]$.
The CRC-aided polar codes are constructed by concatenating the information bits with CRC bits to verify the correctness of each path during decoding.
This improves the error-correction performance, and the polar code is defined as $(N, K+C)$, where $C$ is the number of CRC bits.

\subsection{SCL and SCLF Decoding}

\subsubsection{SCL Decoder}
The SCL decoding algorithm improves the performance of SC decoding by preserving $L$ paths to avoid neglecting the correct path.
Let the received vector be $y_1^N$.
The path metric (PM) for the $i$-th bit of path $l$ in \cite{SCL_2} is introduced in order to measure the likelihood of each path being decoded correctly, and is denoted as:
\begin{equation}\label{pm}
    PM^{(i)}_\ell = \begin{cases}
    PM^{(i-1)}_\ell, &\text{if } \hat{u}^{(i)}_\ell=\frac{1}{2}[1-sign(\alpha^{(i)}_\ell)], \\
    PM^{(i-1)}_\ell+|\alpha^{(i)}_\ell|, &\text{otherwise}
    \end{cases}
\end{equation}
where $\alpha^{(i)}_\ell$ is the LLR for the $i$-th bit of path $l$ represented as:
\begin{equation}\label{llr}
    {\alpha^{(i)}_\ell} = \log\frac{\Pr(u_i=0|y^N_1,
    \hat{u}^{i-1}_1)}{\Pr(u_i=1|y^N_1, \hat{u}^{i-1}_1)}, i\in\left\{1, 2, \ldots, N\right\}.
\end{equation}

While decoding at the position of the information bit, the decoding paths will double, considering two possible values, 0 and 1.
Once the number of expanded paths reaches $2L$, there is competition between each path to achieve the $L$ best paths based on the smallest path metric.
Finally, the decoder outputs the path with the minimum path metric.
In addition, the SCL decoder exploiting the CRC detection denoted as CA-SCL decoding, selects the final output path that can further enhance the performance.

\begin{figure}
\centering
\includegraphics[page=1, width=0.45\textwidth]{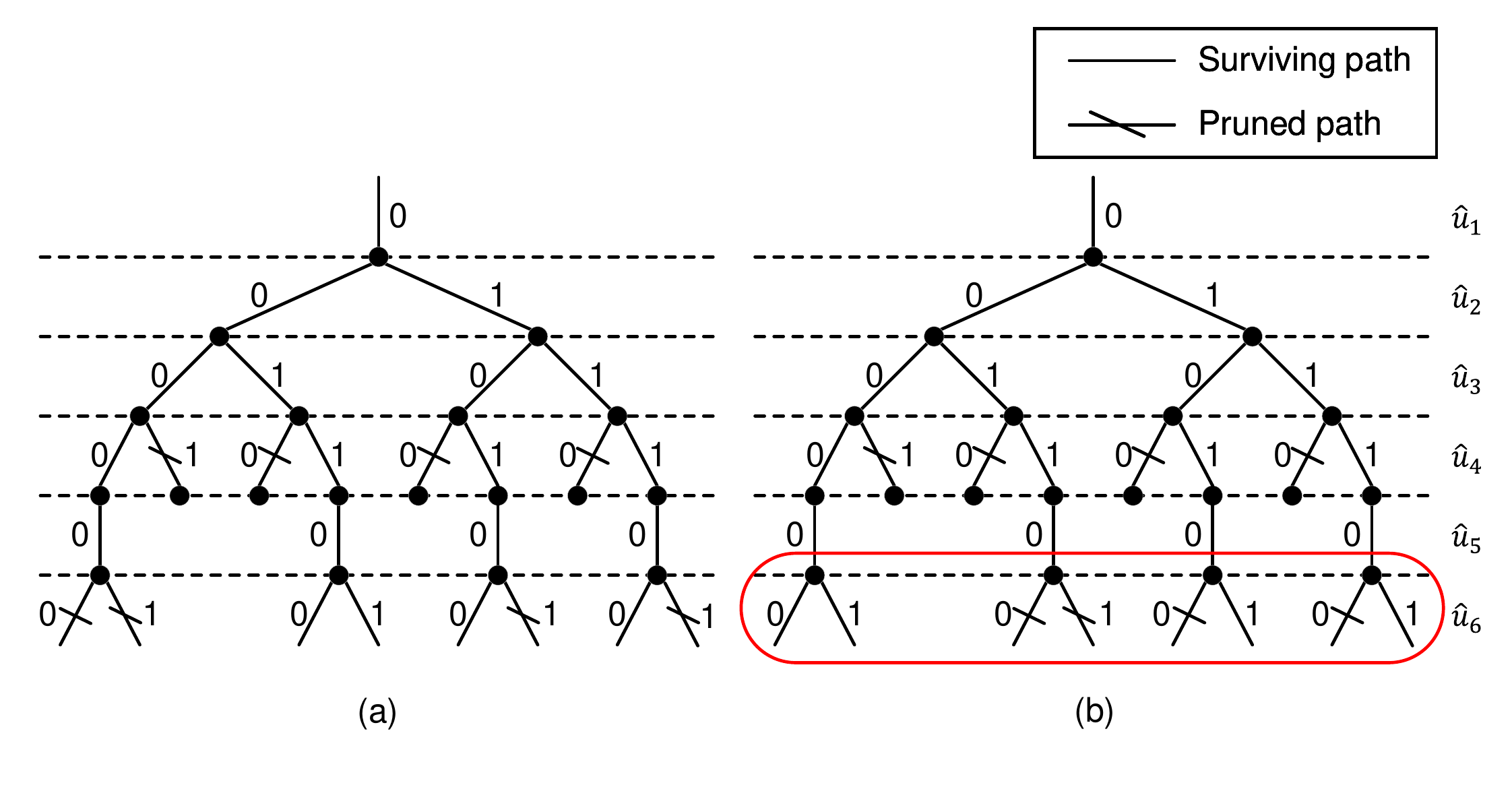}
\caption{Decoding trees for (a) CA-SCL decoding, where $L=4$, and (b) SCLF decoding with the list-flipping procedure at $u_6$.}
\label{fig:flipping_precedure}
\end{figure}

\subsubsection{SCLF Decoder}
If the initial CA-SCL decoding fails the CRC detection, the SCLF decoding method is proposed in order to enhance the performance of the CA-SCL decoding. It will conduct the flipping procedure at the bit position where the correct path may be eliminated during the SCL decoding attempt.
The decoding process continues until the CRC detection is successful or until the maximum number of re-decoding attempts $T$ is reached.
Two kinds of flipping procedures exist, bit-flipping \cite{CASCLF_basic_1}, and list-flipping \cite{CASCLF_basic_2, CASCLF_multi_bit_1, CASCLF_multi_bit_2, CASCLF_ml_1}.
Since the bit-flipping procedure cannot accurately preserve the correct path from the initial CA-SCL decoding, the list-flipping procedure is commonly used.

Fig.~\ref{fig:flipping_precedure}(a) shows an example of CA-SCL decoding, where $L=4$.
Fig.~\ref{fig:flipping_precedure}(b) shows an example of the list-flipping procedure conducted at $u_6$, where the paths eliminated during the initial CA-SCL decoding attempt are preserved, and the preserved paths are simultaneously eliminated.
This guarantees that each of the preserved paths is different from those preserved during the initial CA-SCL decoding attempt, resulting in the correct path being preserved in a subsequent decoding attempt.
In contrast, once we conduct the list-flipping procedure at the bit position where the correct path is already preserved during the initial CA-SCL decoding, it may result in additional incorrect decoded bits during the decoding trajectory.
Thus, it is crucial that we determine the bit position where the correct path is eliminated.
Many previous works have proposed methods for dynamically constructing the flipping set comprised of bit positions according to the metric calculated following the failure of each decoding.

The authors of \cite{CASCLF_multi_bit_1} proposed a generalized SCLF decoding method that is able to flip multiple times during each decoding attempt.
Let $\mathcal{B}$ denote a set comprising the first $\log^L_2$ information bit indices.
The expanded paths at the bit positions in $\mathcal{B}$ are all preserved.
The error metric $E_i$ for the $i$-th bit, $i\in\mathcal{A}\backslash\mathcal{B}$, is calculated from the ratio of the path metrics of the preserved paths to the path metrics of the discarded paths \cite{CASCLF_multi_bit_1} as:
\begin{equation}\label{eq:Ei_1}
    E_i=\log\frac{\sum^{L}_{\ell=1}e^{-PM^{(i)}_\ell}}{\left(\sum^{L}_{\ell=1}e^{-PM^{(i)}_{\ell+L}}\right)^{\beta}}, i\in\mathcal{A}\backslash\mathcal{B},
\end{equation}
where $\beta$ is set to 1.2 according to the Monte-Carlo simulation.
The metric $E_i$ is used to determine the flipping bit position.
When the value $E_i$ is closer to 0, the correct path is more difficult to be determined and preserved. Similarly, the smallest value $E_i$ means the highest probability of conducting list-flipping at bit position $i$.
However, the logarithmic and exponential computations make it problematic when calculating the metric of (\ref{eq:Ei_1}).

The error metric is simplified using the Taylor series approximation as \cite{CASCLF_ml_1}:
\begin{equation}\label{eq:Ei_2}
    E_i=\sum^{L}_{\ell=1}{PM}^i_{\ell+L}-\sum^{L}_{\ell=1}{PM}^i_{\ell}, i\in\mathcal{A}\backslash\mathcal{B},
\end{equation}
and (\ref{eq:Ei_2}) is further approximated as \cite{SCLF_NRB}:
\begin{equation}\label{eq:Ei_3}
    E_i={PM}^i_{L+1}-{PM}^i_1, i\in\mathcal{A}\backslash\mathcal{B},
\end{equation}
where ${PM}^i_1$ and ${PM}^i_{L+1}$ represent the smallest and the $(L+1)$-th smallest value for the path metrics at bit position $i$.
From our simulation results, we observe that the SCLF decoding schemes that have a metric of (\ref{eq:Ei_2}) and (\ref{eq:Ei_3}) are able to achieve almost the same performance.

\subsection{LSTM Networks}


The LSTM network, first introduced by Hochreiter and Schmidhuber \cite{LSTM}, has become popular in recent years.
A highly beneficial feature of the LSTM network makes it possible to store the previous information, thereby providing a high dependency between each basic cell, meaning that it has been suitable for identifying erroneous bit positions in many earlier works, such as \cite{CASCLF_ml_1, CASCLF_ml_2}.

In \cite{CASCLF_ml_1}, the error metric computed from (\ref{eq:Ei_2}) is used as the single feature to train the LSTM model.
However, as the code length increases, it becomes difficult to train the model.
To address this issue, the concept of the critical set \cite{SCF_multi_bit_1} is applied so as to reduce the number of neurons in the output layer, but this reduces the prediction accuracy.
Another approach proposed in \cite{CASCLF_ml_2} applies the LSTM model to each of the $L$ candidate paths to correct the path that fails the CRC detection.
For each path, the network is used to detect the error pattern of the information sub-block and flip it.

\section{Proposed Stacked LSTM Multi-Feature Model}

\begin{figure}
\centering
\includegraphics[page=9, width=0.30\textwidth]{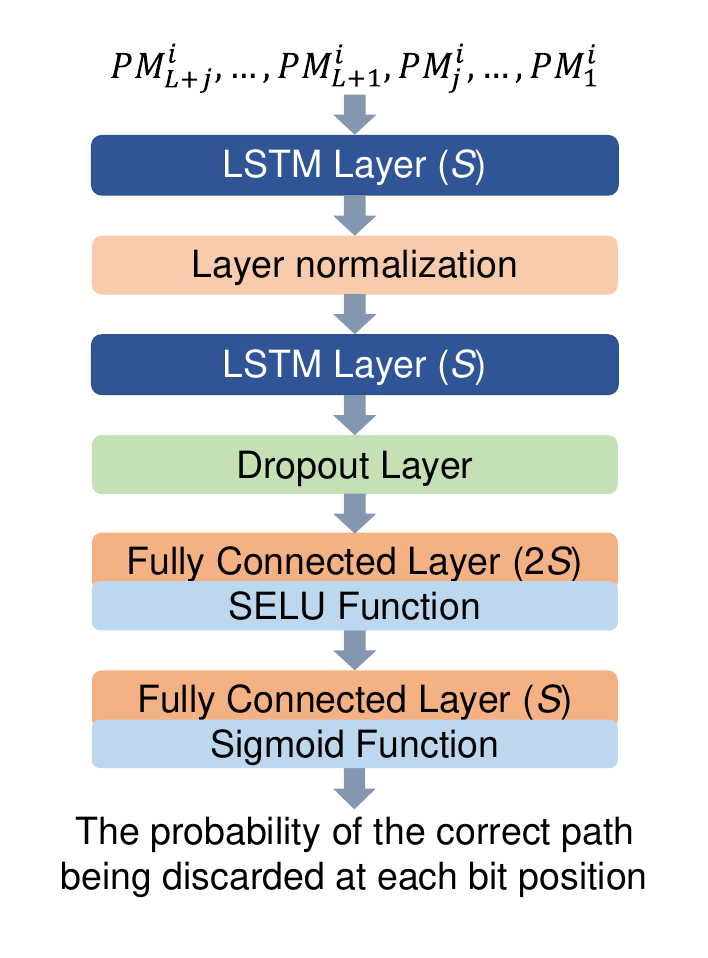}
\caption{Prediction architecture of flipping positions based on the stacked LSTM model.}
\label{fig:architecture_flipping_model}
\end{figure}

\begin{figure}
\centering
\includegraphics[page=29, width=0.45\textwidth]{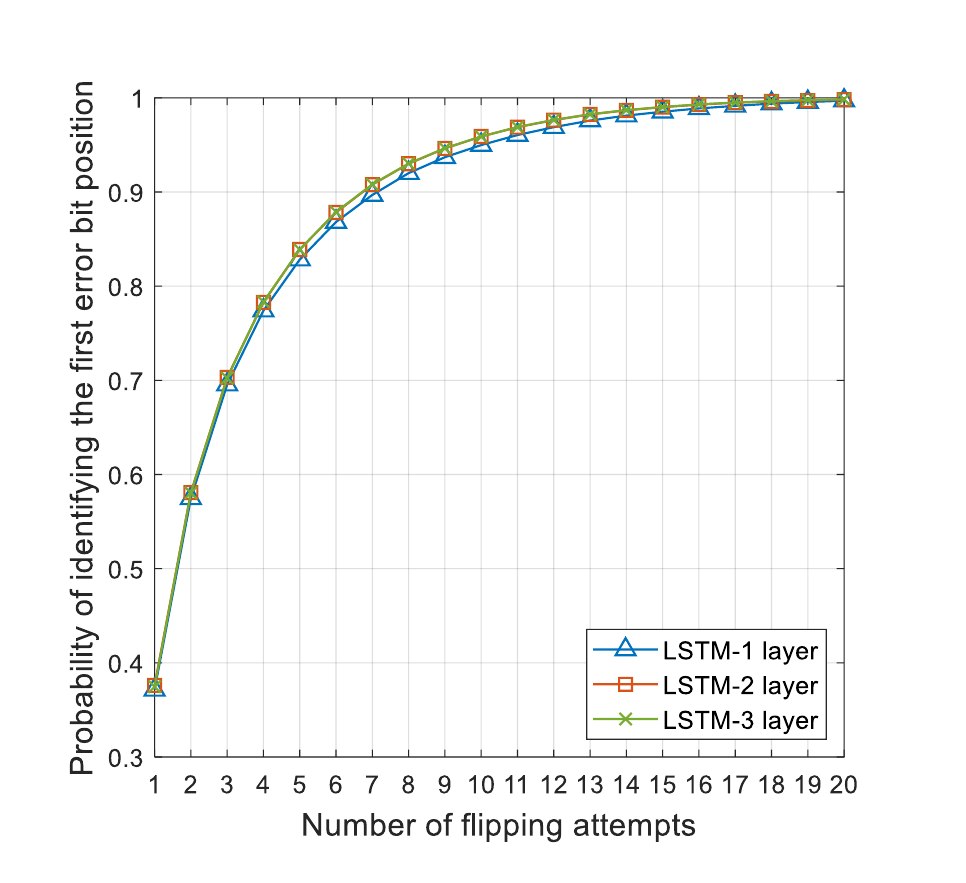}
\caption{\slu{Probability of identifying the position of the first erroneous bit using a different number of LSTM layers for (128, 56+8) polar codes at SNR=2 dB.}}
\label{fig:Cdf_flip1_LSTM_layer}
\end{figure}

\subsection{Stacked LSTM Flipping Model}
\subsubsection{Network Design}
From the conclusion of the experiment conducted in \cite{Stacked_LSTM}, the depth of the LSTM is more critical than the number of LSTM cells within the layer.
In other words, the increase in layer size does not significantly improve the accuracy of the model.
Hence, we adopt the stacked LSTM model of two LSTM layers to assist with the flipping during SCLF decoding and the continue-flipping check.
We propose a stacked LSTM model rather than the metric-based sorting method used in \cite{CASCLF_basic_2, CASCLF_multi_bit_1, CASCLF_multi_bit_2} to predict the erroneous bit position.
The architecture of the model depicted in Fig.~\ref{fig:architecture_flipping_model} consists of two LSTM layers, one dropout layer, and two fully-connected layers, where the last fully-connected layer is used as the output layer.
\slu{
Fig.~\ref{fig:Cdf_flip1_LSTM_layer} illustrates the impact of varying the number of LSTM layers on model accuracy.
The model with a single LSTM layer performs inferior to the other two models consisting of two and three LSTM layers respectively.
Given that the models with two and three LSTM layers demonstrate the same prediction accuracy, we have chosen a model with two LSTM layers for our flipping model, in order to balance accuracy and model complexity.}
For the flipping model, the size of two LSTM layers is $S$, the size of the fully connected layer is expanded to $2S$, and the size of the output layer reduces to $S$, where $S=K+C-\log_2(L)$.

The error metric and the absolute value of the LLR are used to train the LSTM model to identify the bit position conducting the list-flipping procedure in previous works \cite{CASCLF_ml_1, CASCLF_ml_2}.
\slu{However, these studies only exploit a single feature and are yet to maximize the advantages of the LSTM model.}
In this work, we propose multiple features for prediction.
We use the path metric from each path following the CA-SCL decoding process with a list size $L$ as the input for our model.
We select path metrics as the features from both the discarded paths and the preserved paths, represented as follows: \begin{align}\label{eq:proposed_features}
     {PM}^i_{L+j}, \dots, {PM}^i_{L+1}, {PM}^i_j, \dots, {PM}^i_1
\end{align}
where $i\in\mathcal{A}\backslash\mathcal{B}$,  $1 \leq j \leq L$ and ${PM}^i_{L+j}, \dots, {PM}^i_{L+1}$ are the path metrics for the discarded paths and ${PM}^i_j, \dots, {PM}^i_1$ are those for the preserved paths.
The output is the probability of the correct path being discarded at each bit position.


\slu{
Different activation functions of the fully-connected layer and output layer will affect the accuracy of the flipping model.
In our work, we discuss the accuracy of the model designed by the combination of the fully-connected layer and output layer with different activation functions as shown in Table~\ref{tab:flipping_model_activation}. 
According to Table~\ref{tab:flipping_model_activation}, the flipping model consisting of the fully-connected layer containing the scaled exponential linear units (SELU) function \cite{SELU} and the output layer that includes the sigmoid function is able to achieve the best prediction accuracy.
}

The loss function in the flipping model is categorical cross-entropy widely used in multi-class classifiers, which is computed as:
\begin{equation}\label{eq:los_function}
    Loss=-\sum^S_{s=1}y_s\log(\hat{y}_s),
\end{equation}
where $y_s$ denotes the true label, and $\hat{y}_s$ is the prediction value.

Two stacked LSTM flipping models are trained for the proposed SCLF decoding algorithm.
When the initial CA-SCL decoding fails the CRC detection, the proposed features of (\ref{eq:proposed_features}) are input into the stacked LSTM flipping model.
The model is trained to identify the position of the first erroneous bit, named the \textit{stacked LSTM flip-1 model}.

While the first list-flipping attempt by the SCLF decoding algorithm fails the CRC detection, the path metrics from this attempt are input to the stacked LSTM flipping model, named the \textit{stacked LSTM flip-2 model}.
This trained stacked LSTM flip-2 model can be employed for identifying the position of the second erroneous bit.

\begin{table}
\centering
\caption{\slu{Top-5 accuracy of flipping models with different activation functions considered at SNR=2 dB.}}
\label{tab:flipping_model_activation}
\setlength\extrarowheight{5pt}
\setlength{\tabcolsep}{12pt}
\begin{tabular}{|c|c|c|}
\hline
Fully connected layer{\textbackslash}Output layer        & Softmax        & Sigmoid               \\ \hline
Tanh                                                     & 0.83824        & 0.83833               \\ \hline
ReLU                                                     & 0.83744        & 0.83602               \\ \hline
SELU                                                     & 0.83865        & \textbf{0.83895}      \\ \hline
Sigmoid                                                  & 0.83795        & 0.8385                \\ \hline
\end{tabular}
\end{table}

\subsubsection{Training Set}

\begin{figure}
\centering
\includegraphics[page=30, width=0.45\textwidth]{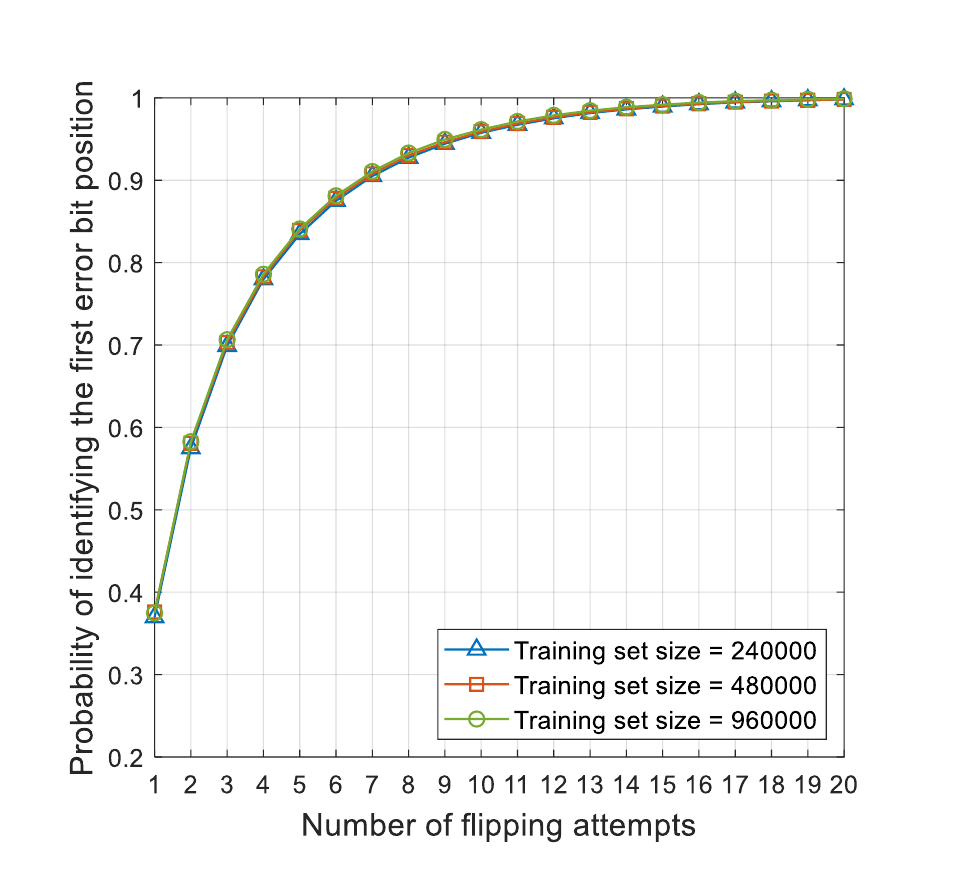}
\caption{\slu{Probability of identifying the position of the first erroneous bit for (128, 56+8) polar codes at SNR=2 dB when using models obtained using  different training sizes.}}
\label{fig:Cdf_flip1_different_training_size}
\end{figure}

Since the noise from the communication system varies, it is difficult to design a general model that doesn't lose performance at each SNR.
Therefore, most work has trained different models that included weights for each SNR.
However, it is difficult to adopt different models in real communication systems.
Therefore, we use an adaptive data set that contains received signals simultaneously suffering from low to high SNR scenarios.
\slu{The data for this work is sourced from SNR scenarios at 1 dB, 1.25 dB, 1.5 dB, 1.75 dB, 2 dB, and 2.25 dB.
In addition, we compared the probabilities of identifying the position of the first erroneous bit using different training dataset sizes of $2.4\times10^5$, $4.8\times10^5$, $9.6\times10^5$ in Fig.~\ref{fig:Cdf_flip1_different_training_size}.
We found that all of them exhibited the same top-5 accuracy probabilities.
Since the training time increases with the size of the dataset, the flipping model was trained using a total of $2.4\times10^5$ training samples and $4.8\times10^4$ validation samples with a training dataset of $4\times10^4$ samples and a validation dataset of $8\times10^3$ samples of each SNR.}
The mini-batch size is 200, the dropout rate for the dropout layer is 0.05, and the optimizer is Adam.

\begin{figure}
\centering
\includegraphics[page=8, width=0.45\textwidth]{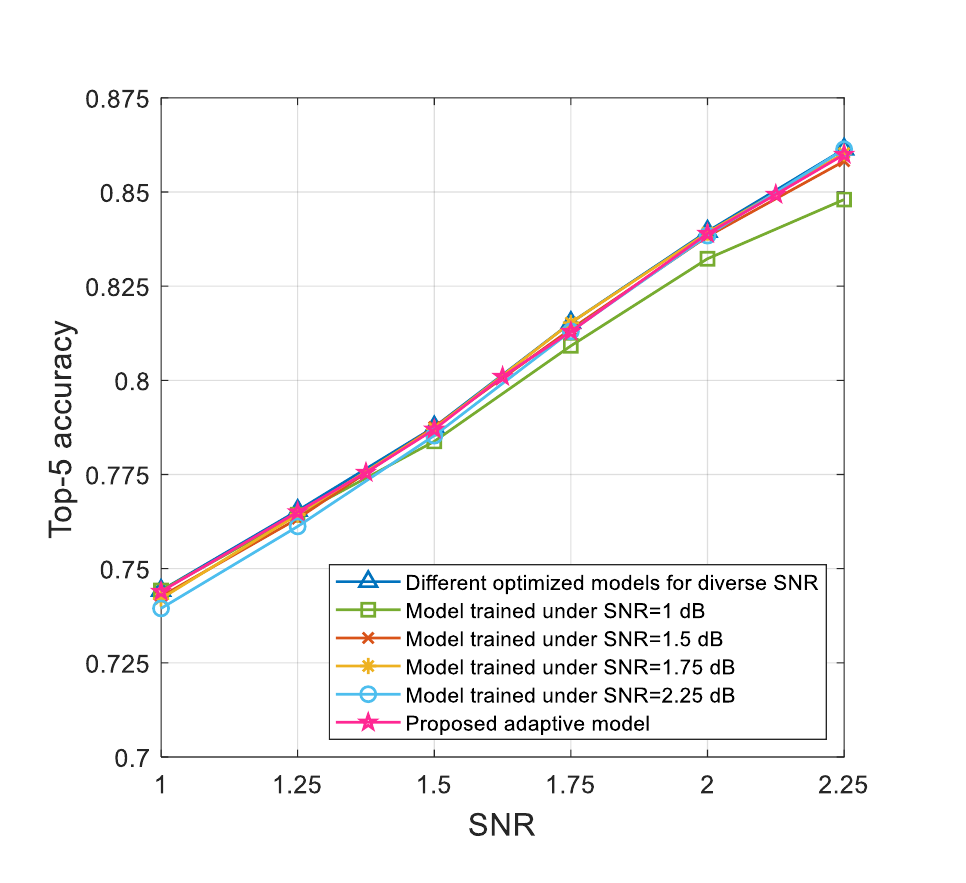}
\caption{Proposed models with the Top-5 accuracy trained under diverse SNRs.}
\label{fig:model_accuracy_for_different_SNRs}
\end{figure}

In this work, the top-5 accuracy is used to evaluate the training model, which means that the top-5 highest probability answers from the model match the expected answer.
Fig.~\ref{fig:model_accuracy_for_different_SNRs} demonstrates the top-5 accuracy for the adaptive data set.
For comparison, the top-5 accuracy for models trained under individual SNRs is also shown.
As observed, the curve for the model trained under SNR=1 dB has the best accuracy at low SNR, but has the worst accuracy for high SNR scenarios.
Furthermore, the curve for the model trained under SNR=2.25 dB shows the best accuracy at a high SNR, but shows the worst for low SNR scenarios.
\slu{
Since the SNR of the communication system will change at any time, the models trained under a specific SNR=1.5 dB and 1.75 dB are not the best choices.
Our adaptive model can alleviate the effect of SNR variations and show good accuracy in the SNR range we simulated due to training with different SNR data.
Not only the SNRs specified in Fig.~\ref{fig:model_accuracy_for_different_SNRs}, our adaptive model also approaches the curve for the different optimized models for diverse SNRs, such as the SNR=1.375 dB, SNR=1.625 dB, and SNR=2.125 dB. It is worth noting that the proposed model works well for other SNRs that the model has never seen before.
}

\subsubsection{Performance Evaluation}
\begin{figure}
\centering
\includegraphics[page=3, width=0.45\textwidth]{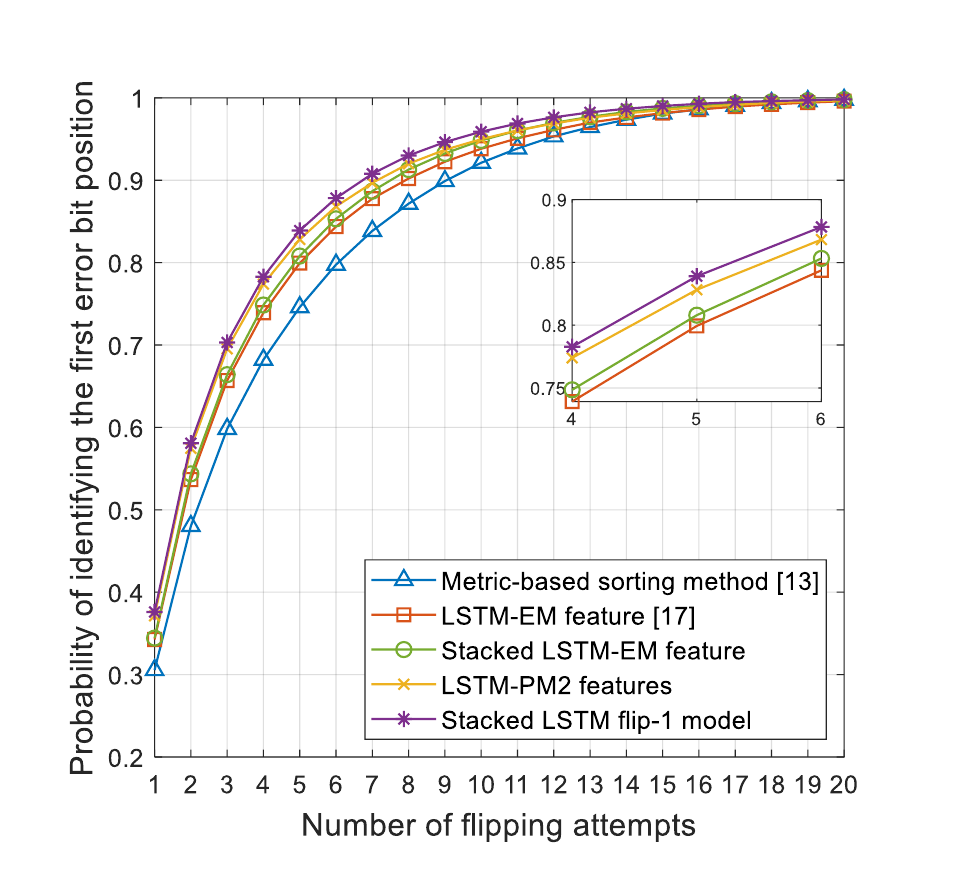}
\caption{Probability of identifying the position of the first erroneous bit using different methods for (128, 56+8) polar codes at SNR=2 dB.}
\label{fig:cdf_flip1_different_methods}
\end{figure}

\begin{figure}
\centering
\includegraphics[page=4, width=0.45\textwidth]{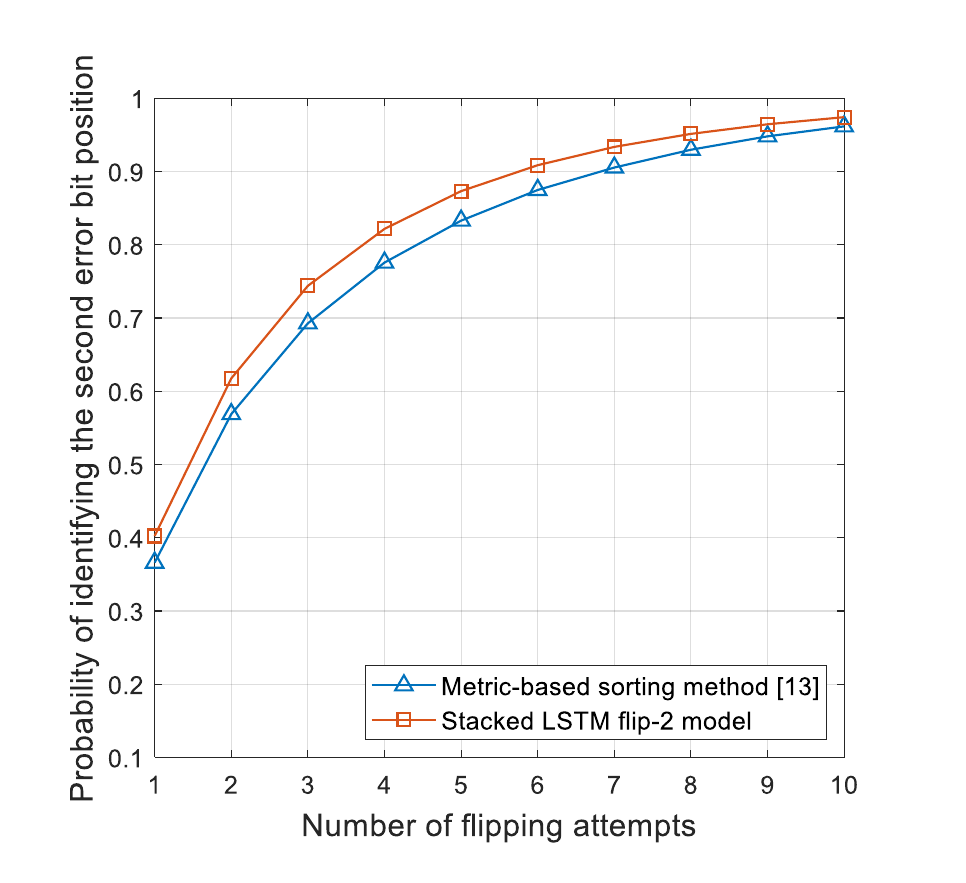}
\caption{Probability of identifying the position of the second erroneous bit using different methods for (128, 56+8) polar codes at SNR=2 dB.}
\label{fig:cdf_flip2_different_methods}
\end{figure}

Fig.~\ref{fig:cdf_flip1_different_methods} illustrates that the stacked LSTM flip-1 model has a higher probability of identifying the position of the first erroneous bit than the other models, such as the metric-based sorting method proposed in \cite{CASCLF_multi_bit_1}, the LSTM model trained using the two proposed features ${PM}^i_{L+1}$ and ${PM}^i_1$ (LSTM-PM2 features model), the LSTM model trained using the error metric calculated from (\ref{eq:Ei_2}) (LSTM-EM feature model), and the stacked LSTM model trained with the error metric from (\ref{eq:Ei_2}) (stacked LSTM-EM feature).
All the simulations were performed for (128, 56+8) polar codes at SNR=2 dB.

The performance by the stacked LSTM flip-2 model is compared to that of the metric-based sorting method in identifying the position of the second erroneous bit shown in
Fig.~\ref{fig:cdf_flip2_different_methods}.
It demonstrates that the performance by our stacked LSTM flip-2 model is better than the metric-based sorting method in identifying the position of the second erroneous bit.

It is observed in Figs.~\ref{fig:cdf_flip1_different_methods} and \ref{fig:cdf_flip2_different_methods} that there is still room for improvement for the error metric proposed in \cite{CASCLF_multi_bit_1}.
This is because the post-computation process in the original error metric causes a loss of information related to the paths at each bit position.
However, by extracting raw data from the error metric of (\ref{eq:proposed_features}), the dependency between each bit position can be revealed, leading to improved prediction accuracy.

\begin{figure}
\centering
\includegraphics[page=5, width=0.45\textwidth]{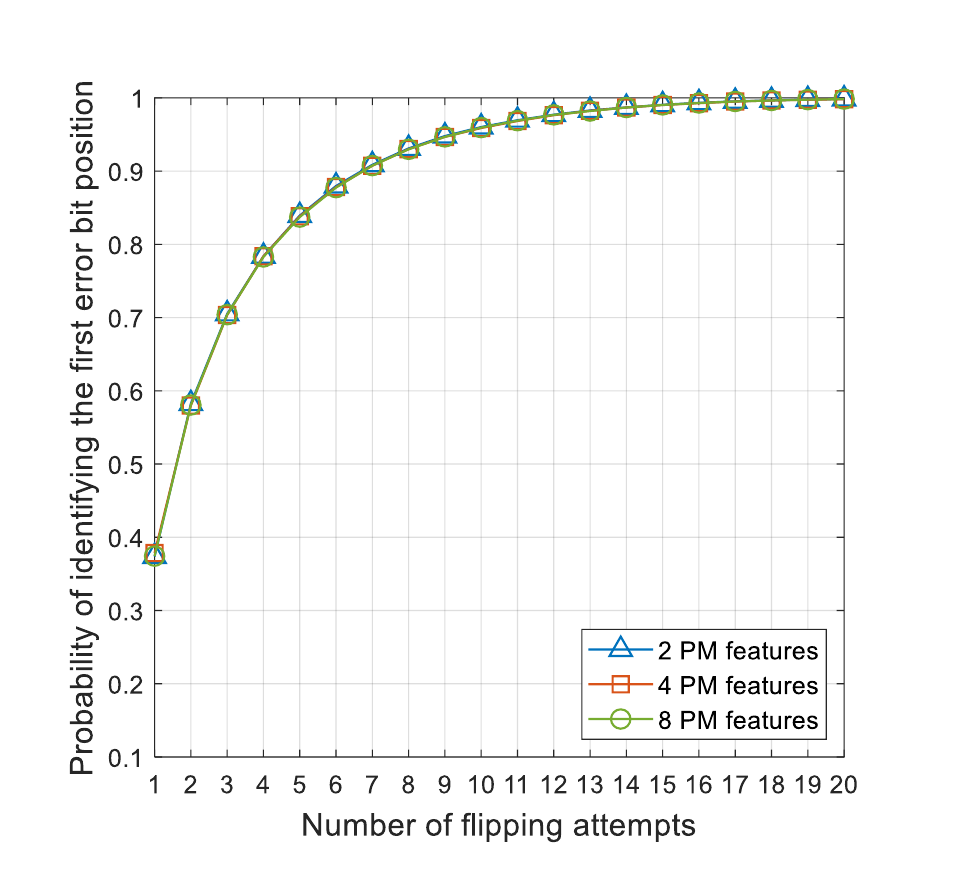}
\caption{Accuracy comparison of the stacked LSTM flip-1 model trained with the different number of path metrics as the features for (128, 56+8) polar codes at SNR=2 dB.}
\label{fig:different_PMs}
\end{figure}

Fig.~\ref{fig:different_PMs} shows a comparison of the prediction accuracy of the models trained using different numbers of path metrics.
Two features consist of ${PM}_{L+1}$ and ${PM}_1$. Four features consist of ${PM}_{L+2}$, ${PM}_{L+1}$, ${PM}_2$, and ${PM}_1$. Eight features consist of ${PM}_{L+4}, \dots, {PM}_{L+1}$, ${PM}_4, \dots, {PM}_1$.
The experiment shows that the number of features is not proportional to the prediction accuracy. Therefore, we selected two features to train our model for low complexity.

\subsection{Stacked LSTM Continue-Flipping Check Model}

\begin{figure}
\centering
\includegraphics[page=10, width=0.30\textwidth]{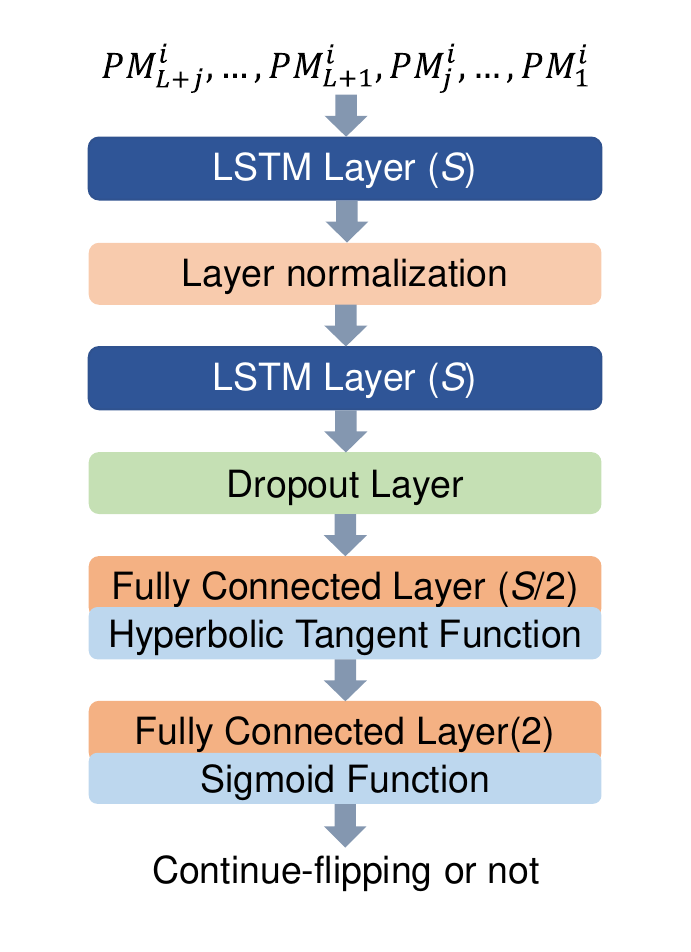}
\caption{Prediction architecture of continue-flipping check result based on the stacked LSTM model.}
\label{fig:architecture_continue-flipping_model}
\end{figure}

\begin{table}
\centering
\caption{\slu{Accuracy of continue-flipping models with different activation functions considered at SNR=2 dB.}}
\label{tab:continue-flipping_model_activation}
\setlength\extrarowheight{5pt}
\setlength{\tabcolsep}{12pt}
\begin{tabular}{|c|c|c|}
\hline
Fully connected layer{\textbackslash}Output layer        & Softmax        & Sigmoid             \\ \hline
Tanh                                                     & 0.84165        & \textbf{0.84175}    \\ \hline
ReLU                                                     & 0.81438        & 0.81387             \\ \hline
SELU                                                     & 0.80859        & 0.84165             \\ \hline
Sigmoid                                                  & 0.80986        & 0.80962             \\ \hline
\end{tabular}
\end{table}

\begin{table}
\centering
\caption{\slu{Accuracy of continue-flipping models with the different number of LSTM layers considered at SNR=2 dB.}}
\label{tab:continue-flipping_different_model}
\setlength\extrarowheight{5pt}
\setlength{\tabcolsep}{24pt}
\begin{tabular}{|c|c|c|}
\hline
Single-layer LSTM       & Stacked LSTM     \\ \hline
0.83825        & \textbf{0.84175}    \\ \hline
\end{tabular}
\end{table}

The stacked LSTM continue-flipping check (CFC) model is designed to determine whether to continue the flipping procedure, which helps to reduce the number of re-decoding attempts.
The input to the model are the features from (\ref{eq:proposed_features}), and the model outputs a binary value to determine whether to continue flipping (1) or not (0).
The architecture for the model, as shown in Fig.~\ref{fig:architecture_continue-flipping_model}, is similar to the flipping model with some variations in terms of the activation function and layer size.

\slu{
As shown in Table~\ref{tab:continue-flipping_model_activation} for the best combination of different activation functions, the stacked LSTM CFC model is designed using a combination of a fully-connected layer with a hyperbolic tangent activation function and a fully-connected layer used as an output layer with a Sigmoid activation function.
In addition, the comparison demonstrated in Table~\ref{tab:continue-flipping_different_model} shows the accuracy of the single-layer LSTM model is inferior to that of the stacked LSTM model with the same activation functions.
The architecture for the continue-flipping check model is designed using two LSTM layers of size $S$, a fully connected layer of size $S/2$, and an output layer of size 2, representing the decision of whether to continue flipping or not.
}

The training data for the continue-flipping check model differs from the flip-2 model.
The training data for the stacked LSTM flip-2 model only includes the features from those cases where the first list-flipping attempt is correct, but fails the CRC detection, while the second list-flipping attempt leads to a successful CRC detection.
In contrast, the training data for the continue-flipping check model includes the features from the first list-flipping cases that fail the CRC detection, including both correct and incorrect cases from the first list-flipping.

The model is trained using a mini-batch size of 500 and a dropout rate of 0.05, with the Adam optimizer and categorical cross-entropy loss function.

\section{Proposed DL-Aided SCLF Decoding Scheme}
This section proposes the DL-aided SCLF-1 and SCLF-2 decoding algorithms based on the trained stacked LSTM models.

\subsection{DL-Aided SCLF-1 Decoding}

\begin{algorithm}
\caption{DL-Aided SCLF-1 Decoding}
\label{alg:DL-SCLF-1 decoding}
	\textbf{Input:} $y^N_1$; ~~~~	\textbf{Output:} $\hat{u}^N_1$
\begin{algorithmic}[1]
\STATE Perform CA-SCL decoding and CRC($\hat{u}^N_1$) fails
\STATE Flipping set $\mathcal{F}$ $\gets$ Stacked LSTM flip-1 model 
    \FOR{$t = 1$ to $T$}
            \STATE $\hat{u}^N_1$ $\gets$ SCLF decoding ($f_t$)
        \IF{CRC($\hat{u}^N_1$) = success}
            \STATE return $\hat{u}^N_1$
        \ENDIF
    \ENDFOR
\end{algorithmic}
\end{algorithm}

The DL-aided SCLF-1 decoding (DL-SCLF-1) algorithm uses the stacked LSTM flip-1 model, outlined in \textbf{Algorithm~\ref{alg:DL-SCLF-1 decoding}}.
The decoding process begins with a CA-SCL decoding process and the simultaneous extraction of the proposed features ${PM}^i_{L+1}$ and ${PM}^i_1$.
If the CA-SCL decoding fails the CRC detection, the SCLF decoding conducts a list-flipping process at a bit position.

In this case, the extracted features are fed into the stacked LSTM flip-1 model, which predicts the probability that the correct path is discarded at each bit position.
The predicted probabilities are sorted in descending order, and the corresponding bit positions are used to create the flipping set $\mathcal{F}=\{f_1,f_2, \dots, f_T\}$, where $f_t$ represents the $t$-th flipping bit position.
Let $T$ be the maximum number of re-decoding attempts.
The bit position $f_t$ is selected during the $t$-th SCLF re-decoding attempt.
The decoding process ends once the maximum number of re-decoding attempts $T$ is reached, or once the CRC detection is passed.

\subsection{DL-Aided SCLF-2 Decoding Algorithms}
To better handle the channel-induced erroneous bit, we also consider decoding algorithms that involve flipping two bit positions.
In this subsection, we consider three SCLF-2 decoding algorithms.
The first and second SCLF-2 decoding algorithms are based on the decoding algorithms discussed in \cite{CASCLF_multi_bit_1} and \cite{CASCLF_ml_3}, respectively, with flipping sets generated by our proposed stacked LSTM models.

\subsubsection{Two-Dimension DL-Aided SCLF-2 Decoding}
The two-dimension DL-aided SCLF-2 decoding (2D-DL-SCLF-2) method described in \textbf{Algorithm~\ref{alg:2D-DL-SCLF-2 decoding}} is based on the decoding algorithm presented in \cite{CASCLF_multi_bit_1} using the proposed stacked LSTM flipping models.

For each re-decoding attempt by the DL-SCLF-1 decoding process, we extract the corresponding features. This means that we will collect features ${PM}^i_{t_1, L+1}$ and ${PM}^i_{t_1,1}$ when we perform list-flipping at bit position $f_{t_1}$.
If the proposed DL-SCLF-1 decoding attempt fails the CRC detection after $T_1$ re-decoding attempts, 2D-DL-SCLF-2 decoding is performed.
For each  $t_1\in \{1,2, \dots, T_2^1\}$, we use the stacked LSTM flip-2 model with features ${PM}^i_{t_1, L+1}$ and ${PM}^i_{t_1,1}$ to generate a flipping set $\mathcal{F}^{(2)}_{t_1}= \{f^{t_1}_{1}, f^{t_1}_{2} \dots, f^{t_1}_{T^2_2} \}$ that contains $T^2_2$ elements representing the second flipping bit position when the first flipping bit position is $f_{t_1}$, where $f_{t_1} \leq f^{t_1}_{1}, f^{t_1}_{2} \dots, f^{t_1}_{T^2_2}$.

In the 2D-DL-SCLF-2 decoding algorithm, we perform two-bit list-flipping using the bit positions $(f_{t_1}, f^{t_1}_{t_2} )$ in the CA-SCL decoding algorithm (\underline{line 6 of \textbf{Algorithm~\ref{alg:2D-DL-SCLF-2 decoding}}}).
The decoding process will continue until either the maximum number of re-decoding attempts $T_1 + T^1_2 \times T^2_2$ is reached or the CRC detection is successful.

\begin{algorithm}
\caption{Two-Dimension DL-Aided SCLF-2 Decoding}
\label{alg:2D-DL-SCLF-2 decoding}
    \textbf{Input:} $y^N_1$;   ~~~~	\textbf{Output:} $\hat{u}^N_1$
\begin{algorithmic}[1]
\STATE Perform DL-SCLF-1 decoding and CRC($\hat{u}^N_1$) fails
 \FOR{$t_1 = 1$ to $T^1_2$}
 \STATE (${PM}^i_{t_1, L+1}, {PM}^i_{t_1,1}$) $\gets$ Extract DL-SCLF-1 decoding  $t_1$-th attempt features
      \STATE Flipping set $\mathcal{F}^{(2)}_{t_1}$ $\gets$ Stacked LSTM flip-2 model (${PM}^i_{t_1, L+1}, {PM}^i_{t_1,1}$)
        \FOR{$t_2 = 1$ to $T^2_2$}
                \STATE $\hat{u}^N_1$ $\gets$ SCLF decoding $(f_{t_1}, f^{t_1}_{t_2})$
                \IF{CRC($\hat{u}^N_1$) = success}    \STATE return $\hat{u}^N_1$
                \ENDIF
       \ENDFOR
\ENDFOR
\end{algorithmic}
\end{algorithm}

\subsubsection{DL-Aided Two-Phase SCLF-2 Decoding}
The DL-aided two-phase SCLF-2 decoding (DL-TP-SCLF-2), described in \textbf{Algorithm~\ref{alg:DL-TP-SCLF-2 decoding}}, is based on the algorithm from \cite{CASCLF_ml_3}, with the only difference being that our stacked LSTM model generates the flipping set and the continue-flipping check result.

In this algorithm, the flipping set $\mathcal{F}=\{f_1,f_2, \dots, f_T\}$ is generated based on the predictions by the stacked LSTM flip-1 model.
Suppose the SCLF decoding result fails the CRC detection at the $t$-th decoding attempt.
In that case, we use the proposed features as input for the continue-flipping check model to determine whether to proceed with the next list-flipping step based on the list-flipping results.
\slu{If the model predicts the result to be continued decoding, an SCLF decoding attempt based on two list-flipping is considered, and the two flipping bit positions are $(f_t, f_{t+1})$  (\underline{line 8 of \textbf{Algorithm~\ref{alg:DL-TP-SCLF-2 decoding}}}).}

In DL-aided two-phase SCLF-2 decoding, the second erroneous bit position is chosen from the flipping set $\mathcal{F}$ generated from the initial CA-SCL decoding failure.
However, the decoding trajectory is altered after the list-flipping at the first erroneous bit position, which leads to an inaccurate selection for the second erroneous bit position.

\begin{algorithm}
\caption{DL-Aided Two-Phase SCLF-2 Decoding}
\label{alg:DL-TP-SCLF-2 decoding}
	\textbf{Input:} $y^N_1$;  ~~~~	\textbf{Output:} $\hat{u}^N_1$
\begin{algorithmic}[1]
\STATE Perform  CA-SCL decoding and CRC($\hat{u}^N_1$) fails
\STATE Flipping set $\mathcal{F}$ $\gets$ Stacked LSTM flip-1 model
\FOR{$t = 1$ to $T$}
 \STATE $\hat{u}^N_1$ $\gets$ SCLF decoding ($f_t$)
\IF{CRC($\hat{u}^N_1$) = failure}
  \STATE $C_f$ $\gets$Stacked LSTM continue-flipping check model 
        \IF{$C_f = 1$}
            \STATE $\hat{u}^N_1$ $\gets$ SCLF decoding $(f_t, f_{t+1})$
             \IF{CRC($\hat{u}^N_1$) = success}    \STATE return $\hat{u}^N_1$
                \ENDIF
     \ENDIF
 \ENDIF
\ENDFOR
\end{algorithmic}
\end{algorithm}

\begin{algorithm}
\caption{Proposed DL-Aided SCLF-2 Decoding}
\label{alg:DL-SCLF-2 decoding:proposed method}
	\textbf{Input:} $y^N_1$;  ~~~~	\textbf{Output:} $\hat{u}^N_1$
\begin{algorithmic}[1]
\STATE Perform CA-SCL decoding and CRC($\hat{u}^N_1$) fails
\STATE Flipping set $\mathcal{F}$ $\gets$ Stacked LSTM flip-1 model 
\FOR{$t = 1$ to $T$}
    \STATE $\hat{u}^N_1$ $\gets$ SCLF decoder ($f_t$)
  \IF{CRC($\hat{u}^N_1$) = failure}
  \STATE $C_f$ $\gets$Stacked LSTM continue-flipping check model 
        \IF{$C_f = 1$}
               \STATE $f^{\star}_{t}$ $\gets$ Stacked LSTM flip-2 model 
               \STATE $\hat{u}^N_1$ $\gets$ SCLF decoding $(f_t, f^{\star}_{t})$
                 \IF{CRC($\hat{u}^N_1$) = success}    \STATE return $\hat{u}^N_1$
                \ENDIF
               \ENDIF
    \ENDIF
\ENDFOR
\end{algorithmic}
\end{algorithm}

\subsubsection{Proposed DL-Aided SCLF-2 Decoding}
Based on the stacked LSTM flip and CFC model, we consider a proposed DL-aided SCLF-2 (proposed DL-SCLF-2) decoding approach in \textbf{Algorithm~\ref{alg:DL-SCLF-2 decoding:proposed method}}.

In \textbf{Algorithm~\ref{alg:DL-SCLF-2 decoding:proposed method}}, the flipping set $\mathcal{F}$ is generated based on the stacked LSTM flip-1 model.
If the result from the CA-SCL decoding, where a single position $f_t$ is flipped, fails the CRC detection, the new features are input into the continue-flipping check model.
\slu{If the stacked LSTM continue-flipping check model predicts to continue decoding, the features are fed into the stacked LSTM flip-2 model to predict the second erroneous bit position.}

Once the second erroneous bit position $f^{\star}_{t}$ is determined, it is used to perform the SCLF decoding process with two bit positions $(f_t, f^{\star}_{t})$ flipped (\underline{line 9 of \textbf{Algorithm~\ref{alg:DL-SCLF-2 decoding:proposed method}}}).
The decoding process will conclude when the result passes the CRC detection or the maximum number of re-decoding attempts $T$ is reached.

\textit{Remarks:} The comparison of the flipping bit position indices and the use of the stacked LSTM flip-2 model and the continue-flipping check (CFC) model for the three DL-aided SCLF-2 decoding algorithms are shown in Table~\ref{tab:SCLF2}.
The first list-flipping in all of the SCLF-2 decoding algorithms is based on the stacked LSTM flip-1 model.
However, the second list-flipping is based on different models.
The 2D-DL-SCLF-2 algorithm is only based on the flip-2 model, and the DL two-phase SCLF-2 algorithm is based on the flip-1 model and the CFC model. Furthermore, the proposed DL-aided SCLF-2 decoding algorithm is based on both the flip-2 model and the CFC model.

\slu{
This divergence in the second list-flipping 
provides a theoretical basis for exploring differences in decoding performance and the number of decoding attempts.
In terms of \textit{decoding performance}, the DL-TP-SCLF-2 algorithm predicts the second erroneous bit from the initial decoding failure. However, the decoding trajectory changes after the first erroneous bit is flipped, which may lead to a loss in performance. In contrast, the 2D-DL-SCLF-2 and DL-SCLF-2 algorithms predict the second erroneous bit using a stacked LSTM flip-2 model, which is based on the new decoding trajectory. This approach might yield better performance compared to the DL-TP-SCLF-2 algorithm.}
\slu{
As for the \textit{number of decoding attempts}, the DL-TP-SCLF-2 and DL-SCLF-2 algorithms use the CFC model, which may result in fewer decoding attempts compared to the 2D-DL-SCLF-2 algorithm. Therefore, the proposed DL-SCLF-2 decoding algorithm, which utilizes both the flip-2 and CFC models, might enhance decoding performance and speed. The experimental results of FER and decoding attempts performances are shown in Section V.}

\slu{While the proposed DL-SCLF-2 decoding algorithm may increase complexity due to the requirement for two models, it is essential to note that these models are trained before decoding. Therefore, the primary increase in complexity occurs during the offline phase. The increase in complexity during the online decoding process is relatively minor.}

\begin{table}
\caption{The comparison of three DL-Aided SCLF-2 Decoding Algorithms}
\label{tab:SCLF2}
\centering
    \begin{tabular}{c|c|c|c}
    \hline
Algorithm             & \begin{tabular}[c]{@{}c@{}}Flipping bit \\ position indexes \end{tabular}
                      & \begin{tabular}[c]{@{}c@{}}Flip-2       \\ model            \end{tabular}
                      & \begin{tabular}[c]{@{}c@{}}CFC          \\ model            \end{tabular} \\    \hline
2D-DL-SCLF-2 (Alg.~2) & $(f_{t_1}, f^{t_1}_{t_2})$ & Yes         & No                             \\    \hline
DL-TP-SCLF-2 (Alg.~3) & $(f_t, f_{t+1})$           & No          & Yes                            \\    \hline
DL-SCLF-2 (Alg.~4)    & $(f_t, f^{\star}_{t})$     & Yes         & Yes                            \\
   \hline
  \end{tabular}
\end{table}

\section{Experimental Results}
In this section, we compare the FER performance and complexity of the proposed SCLF decoding algorithms with other SCLF decoding algorithms presented in \cite{CASCLF_basic_2, CASCLF_multi_bit_1}, and \cite{CASCLF_ml_1}.
Our simulation assumes that polar codes are transmitted over an additive white Gaussian noise (AWGN) channel based on binary phase shift keying (BPSK) modulation.
The (128, 56+8) polar code is a (128, 56) polar code concatenated to a CRC detector using the generator polynomial $g(x)=x^8+x^2+x+1$.
The number of average decoding attempts (ADA) is calculated as follows:
\begin{equation}\label{eq:ADA}
    ADA=\frac{1+\sum\#\ \text{of\ re-decoding\ attempts\ per\ frame}}{\text{Total\ frame}}
\end{equation}
where the initial CA-SCL decoding process is considered as a single decoding attempt.

\begin{figure}
    \centering
    \subfloat[FER performance comparison]{
        \includegraphics[page=15, width=0.45\textwidth]{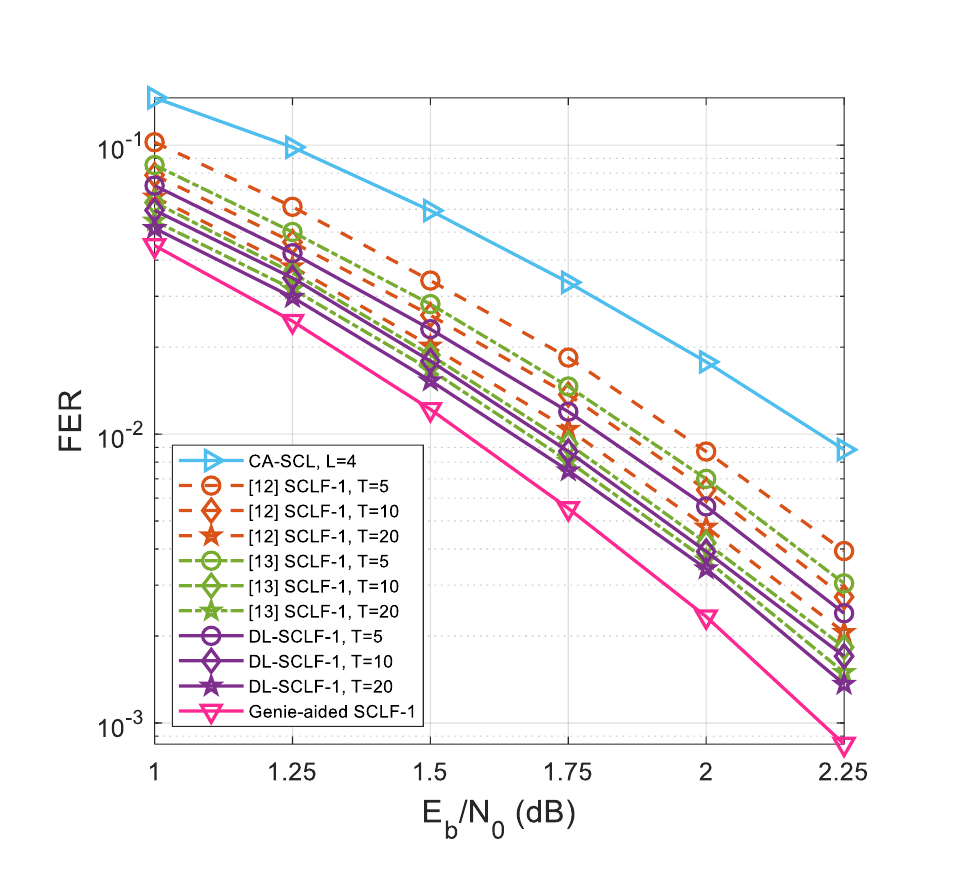}
        \label{fig:FER_SCLF-1 decoding}
    }
    \hfill
    \subfloat[Comparison of the average number of decoding attempts]{
        \includegraphics[page=16, width=0.45\textwidth]{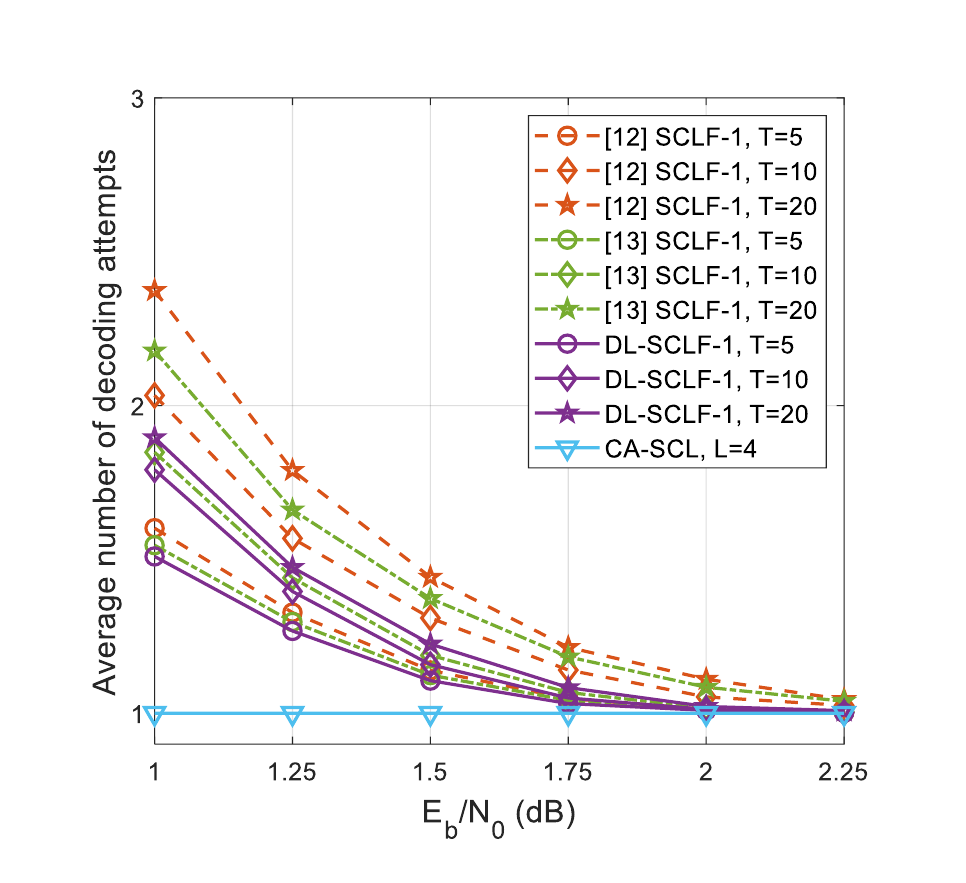}
        \label{fig:ADA_SCLF-1 decoding}
    }
    \caption{Performance comparison for different SCLF-1 decoding schemes for (128, 56+8) polar codes where $L=4$.}
    \label{fig:SCLF-1 decoding}
\end{figure}

\begin{figure}
    \centering
    \subfloat[FER performance comparison]{
        \includegraphics[page=17, width=0.45\textwidth]{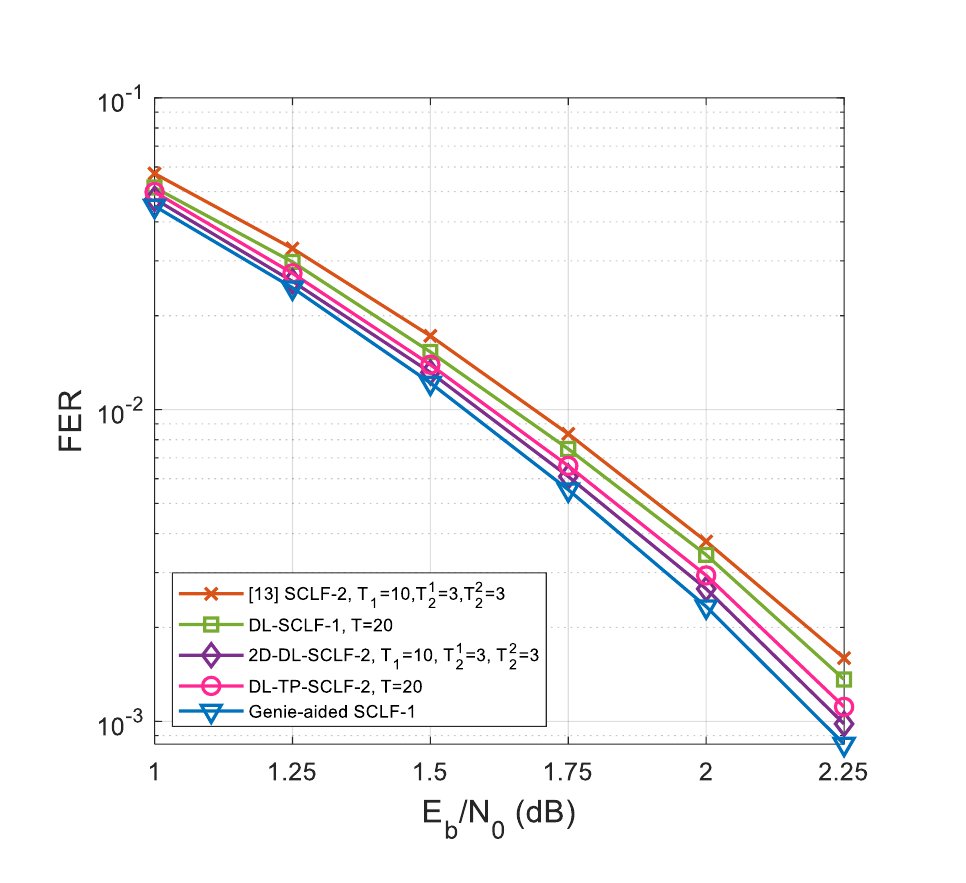}
        \label{fig:FER_SCLF-2 decoding}
    }
    \hfill
    \subfloat[Comparison of the average number of decoding attempts]{
        \includegraphics[page=18, width=0.45\textwidth]{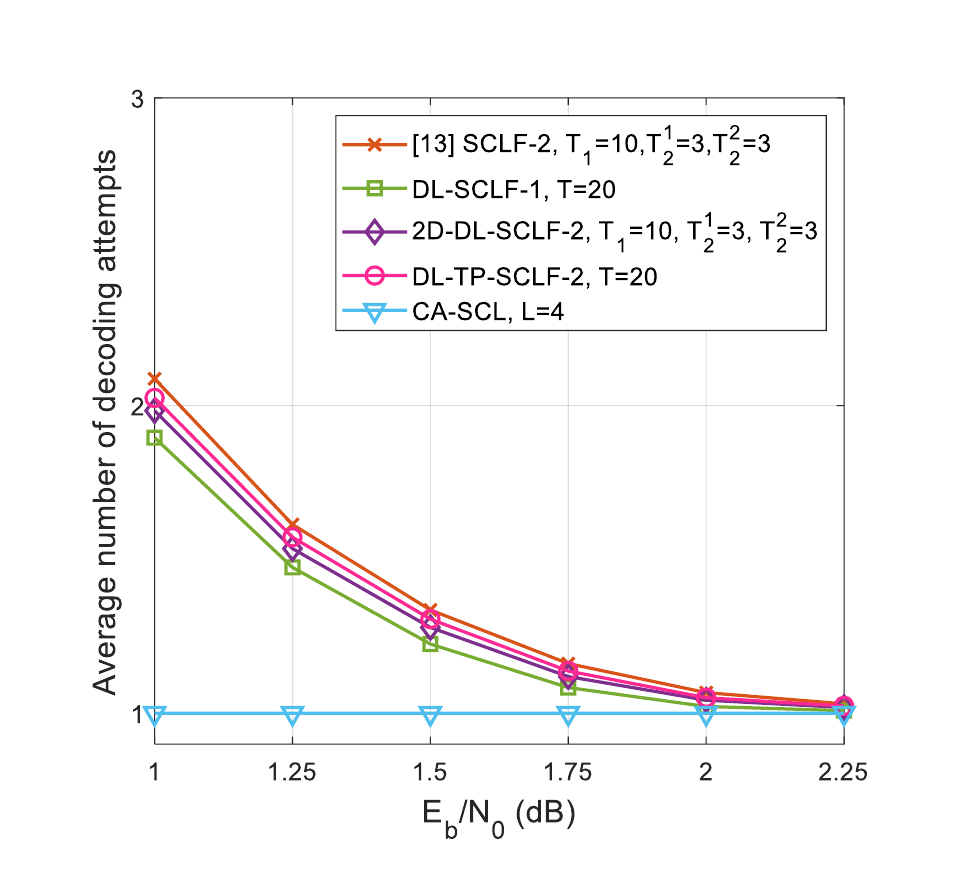}
        \label{fig:ADA_SCLF-2 decoding}
    }
    \caption{Performance comparison for different SCLF-2 decoding schemes for (128, 56+8) polar codes where $L=4$.}
    \label{fig:SCLF-2 decoding}
\end{figure}

\begin{figure}
    \centering
    \subfloat[\slu{FER performance comparison}]{
        \includegraphics[page=19, width=0.45\textwidth]{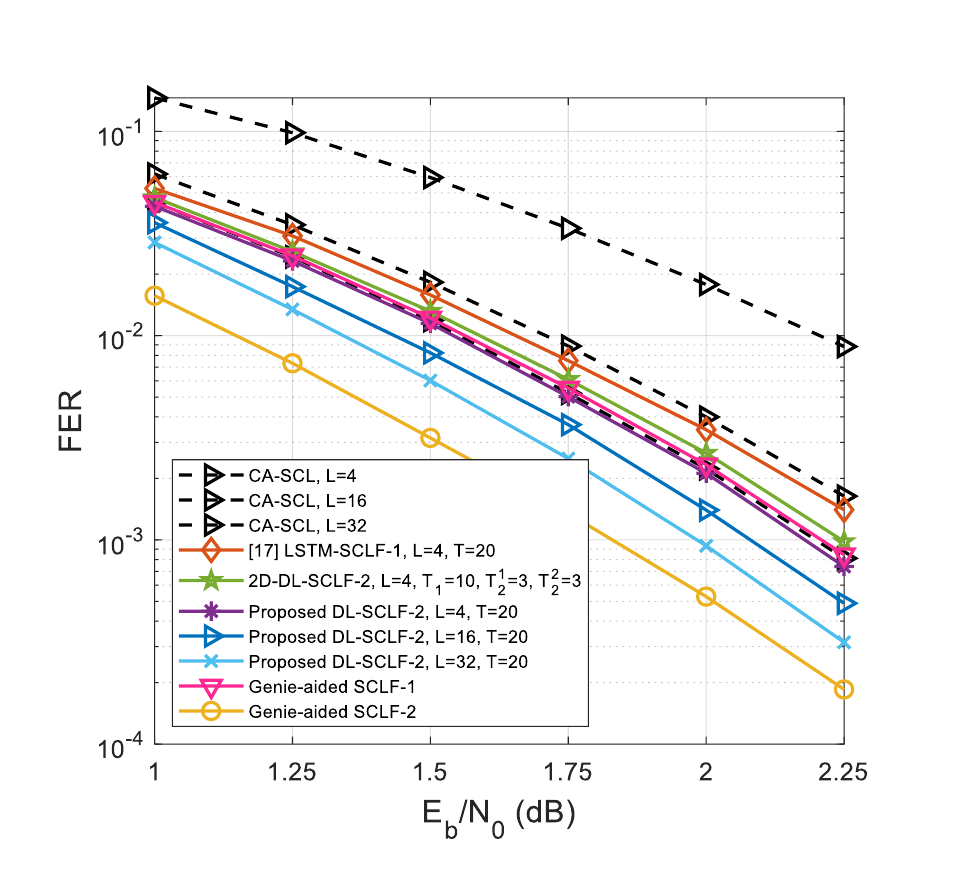}
        \label{fig:FER_proposed SCLF-2 decoding}
    }
    \hfill
    \subfloat[\slu{Comparison of the average number of decoding attempts}]{
        \includegraphics[page=20, width=0.45\textwidth]{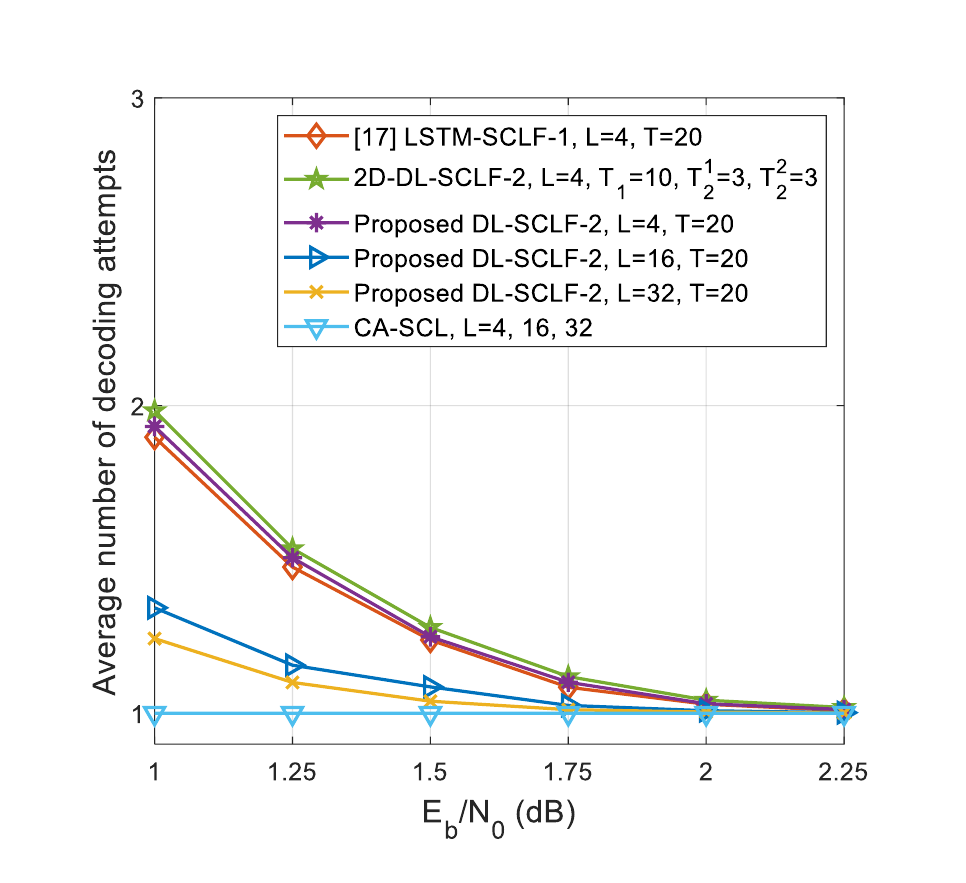}
        \label{fig:ADA_proposed SCLF-2 decoding}
    }
    \caption{Performance comparison of different SCLF-1 and SCLF-2 decoding schemes based on the same total maximum number of re-decoding attempts for (128,56+8) polar codes where $L=4, 16, 32$.}
    \label{fig:proposed SCLF-2 decoding}
\end{figure}

\begin{figure}
    \centering
    \subfloat[FER performance comparison]{
        \includegraphics[page=21, width=0.45\textwidth]{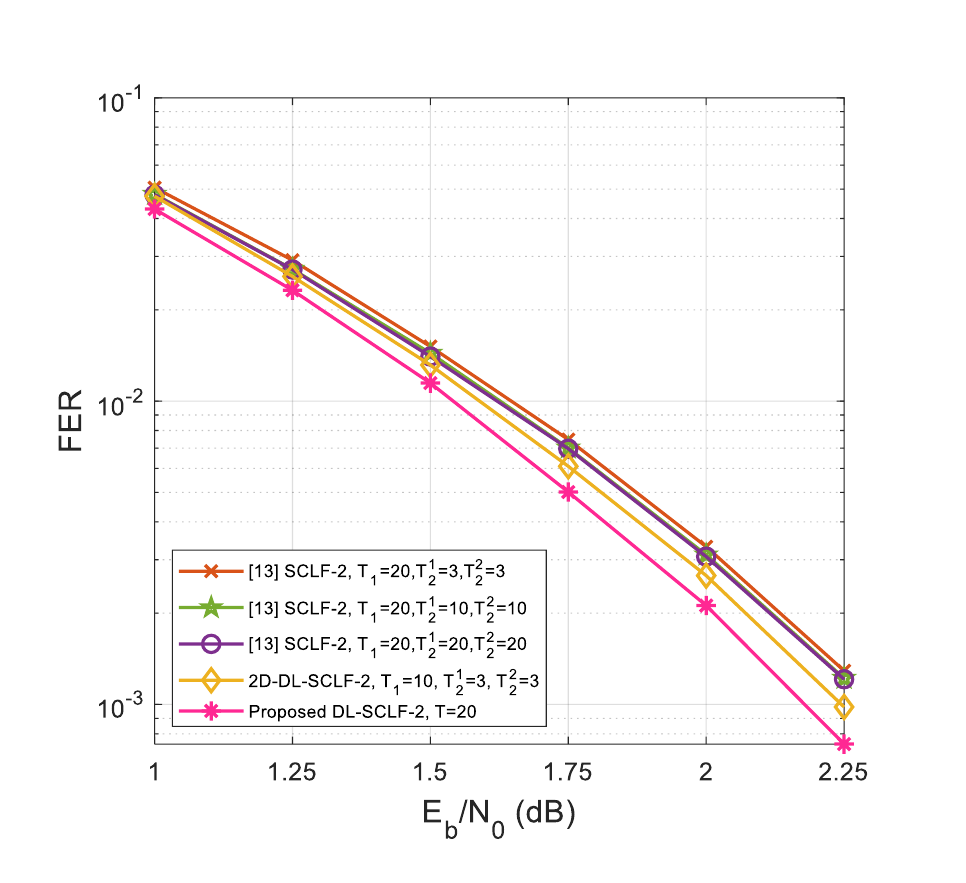}
        \label{fig:FER_diff_iter_proposed DL-SCLF-2 decoding}
    }
    \hfill
    \subfloat[Comparison of the average number of decoding attempts]{
        \includegraphics[page=22, width=0.45\textwidth]{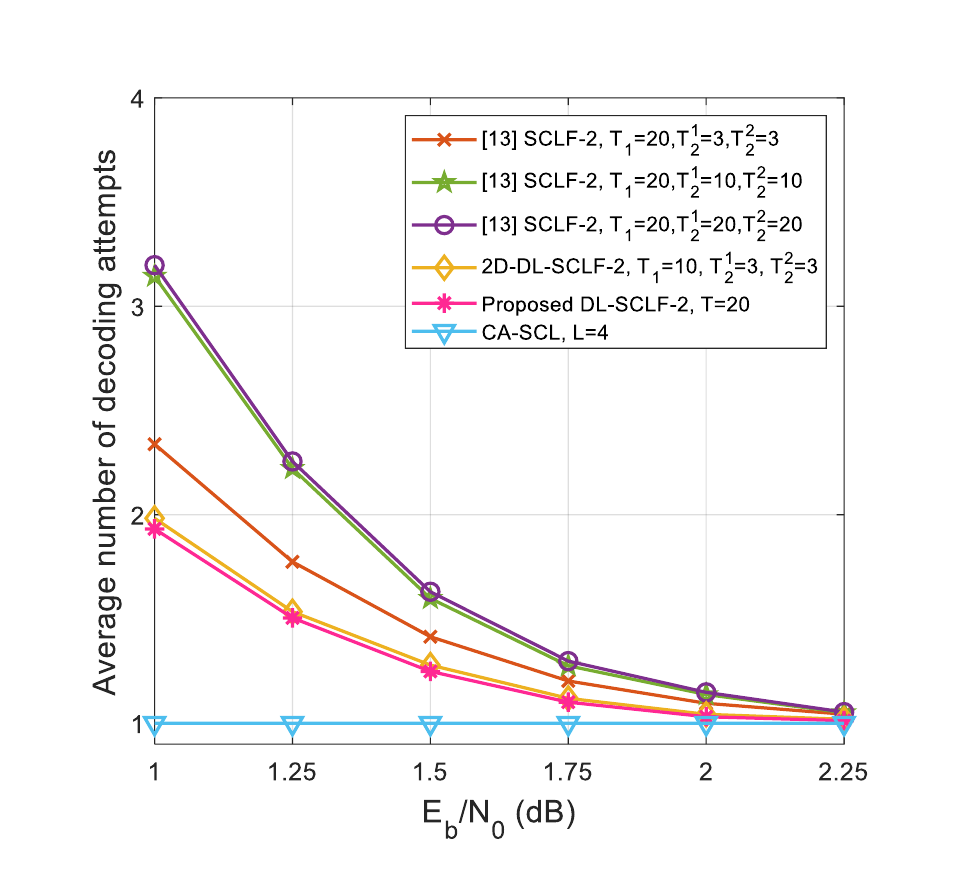}
        \label{fig:ADA_diff_iter_proposed DL-SCLF-2 decoding}
    }
    \caption{Performance comparison of SCLF-2 and DL-SCLF-2 decoding schemes based on different maximum re-decoding attempts for (128, 56+8) polar codes where $L=4$.}
    \label{fig:diff_iter_proposed DL-SCLF-2 decoding}
\end{figure}

\begin{figure}
    \centering
    \subfloat[FER performance comparison]{
        \includegraphics[page=23, width=0.45\textwidth]{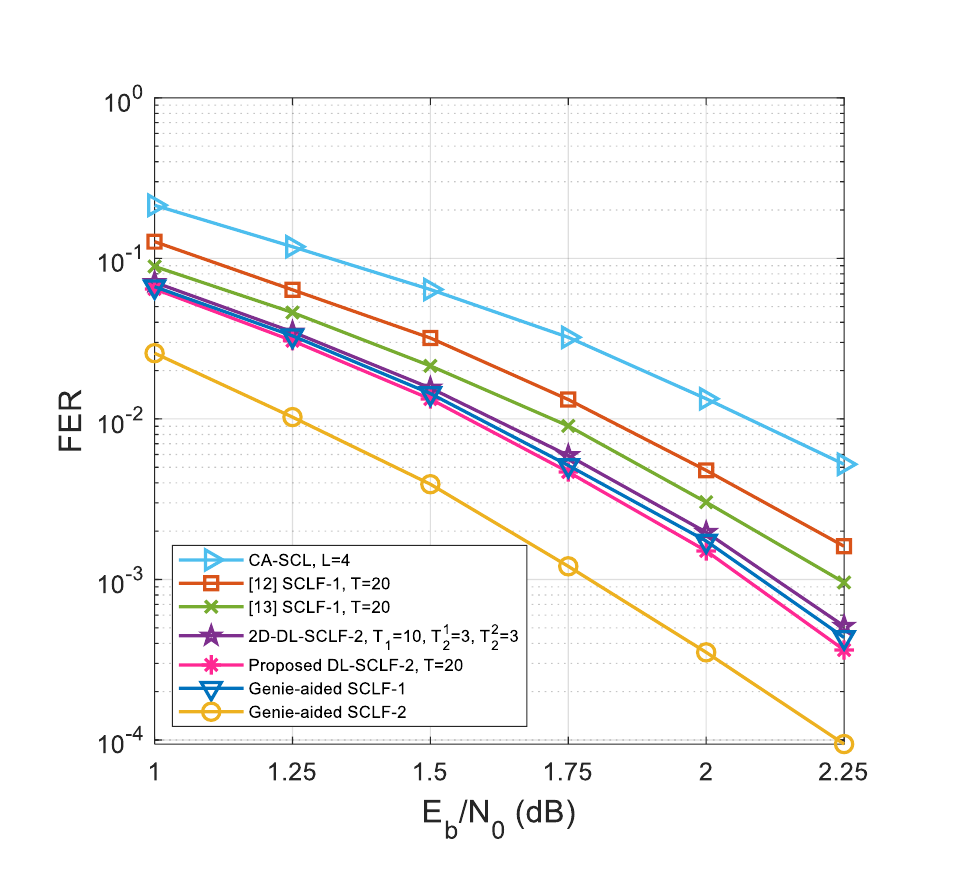}
        \label{fig:FER_N256}
    }
    \hfill
    \subfloat[Comparison of the average number of decoding attempts]{
        \includegraphics[page=24, width=0.45\textwidth]{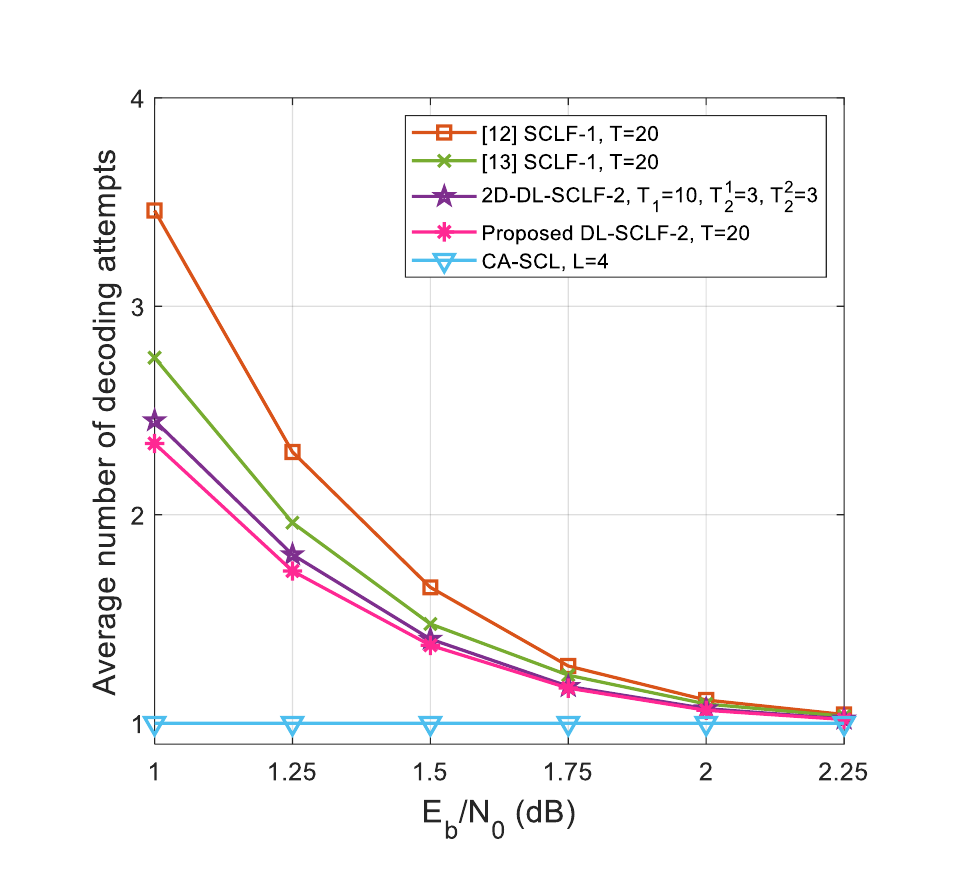}
        \label{fig:ADA_N256}
    }
    \caption{Performance comparison for (256, 120+8) polar codes where $L=4$.}
    \label{fig:N256}
\end{figure}

\begin{figure}
    \centering
    \subfloat[FER performance comparison]{
        \includegraphics[page=25, width=0.45\textwidth]{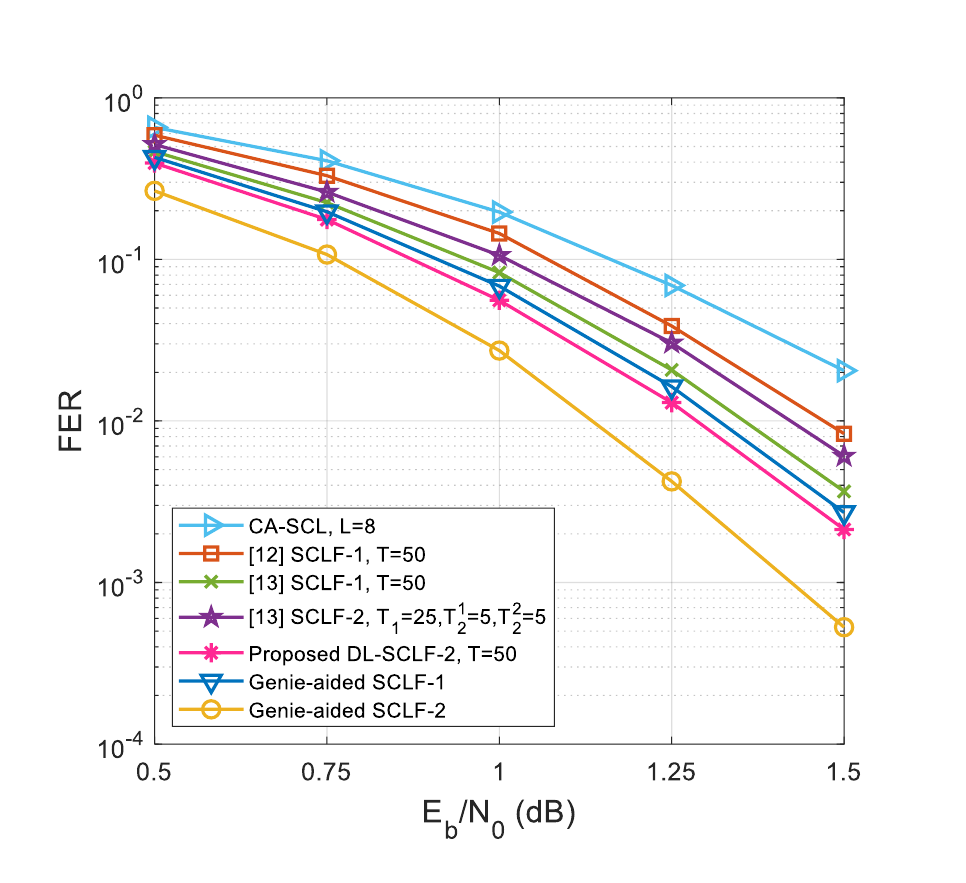}
        \label{fig:FER_N1024}
    }
    \hfill
    \subfloat[Comparison of the average number of decoding attempts]{
        \includegraphics[page=26, width=0.45\textwidth]{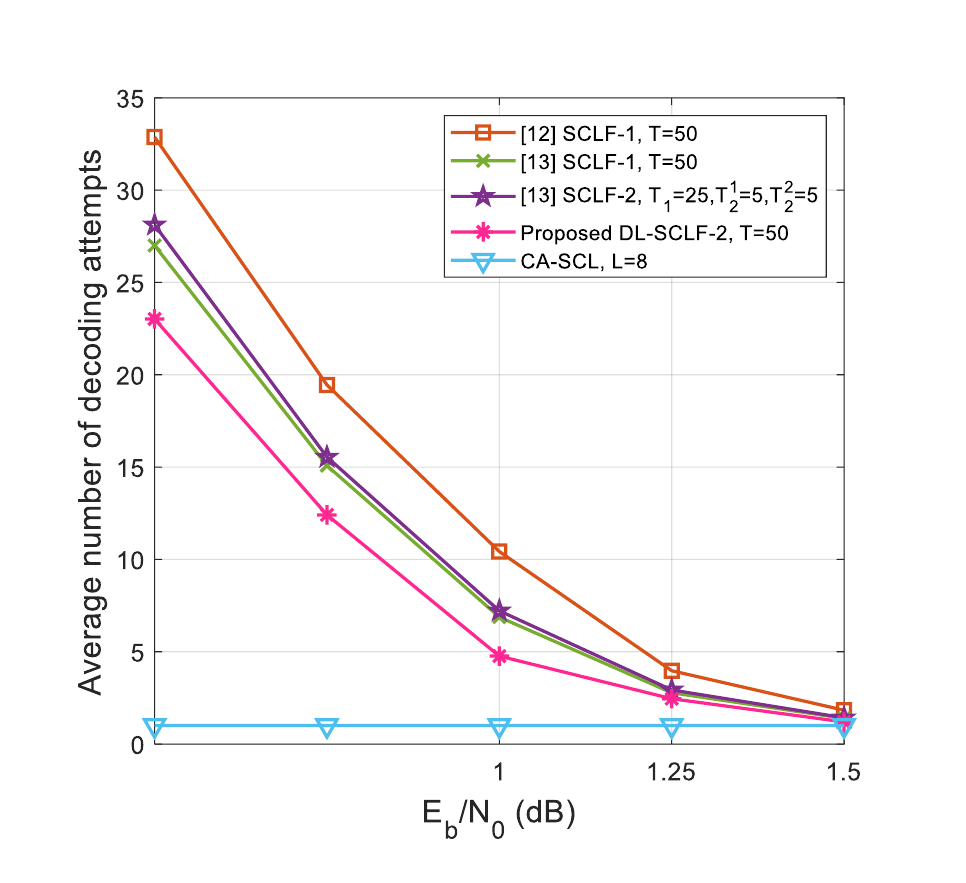}
        \label{fig:ADA_N1024}
    }
    \caption{Performance comparison  for (1024, 496+16) polar codes where $L=8$.}
    \label{fig:N1024}
\end{figure}

\begin{figure}
    \centering
    \subfloat[FER performance comparison]{
        \includegraphics[page=27, width=0.45\textwidth]{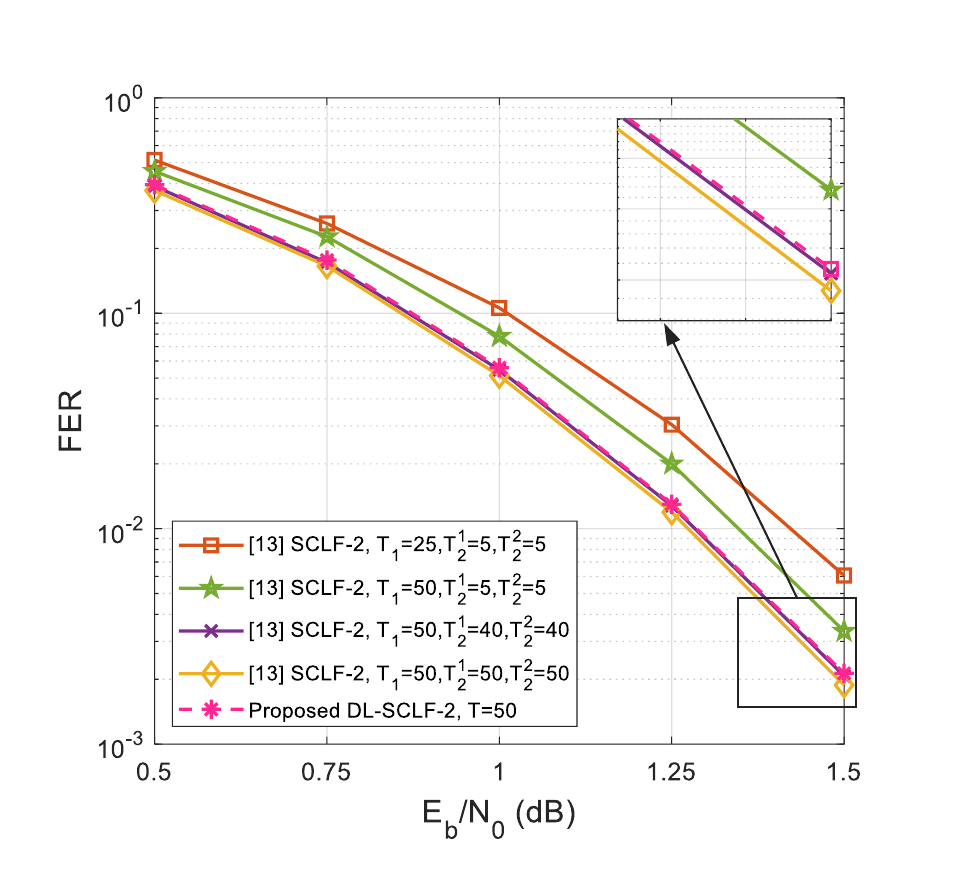}
        \label{fig:FER_diff_iter_N1024}
    }
    \hfill
    \subfloat[Comparison of the average number of decoding attempts]{
        \includegraphics[page=28, width=0.45\textwidth]{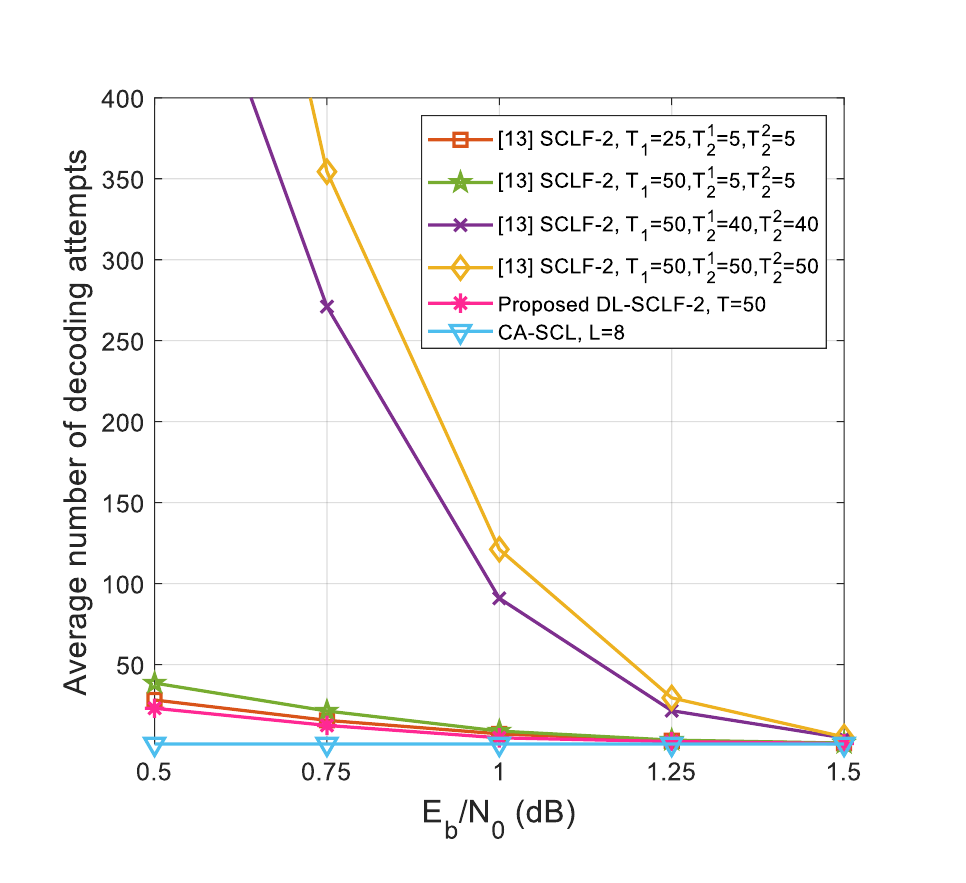}
        \label{fig:ADA_diff_iter_N1024}
    }
    \caption{Performance comparison for (1024, 496+16) polar codes where $L=8$.}
    \label{fig:diff_iter_N1024}
\end{figure}

\begin{figure}
    \centering
    \subfloat[\slu{FER performance comparison}]{
        \includegraphics[page=31, width=0.45\textwidth]{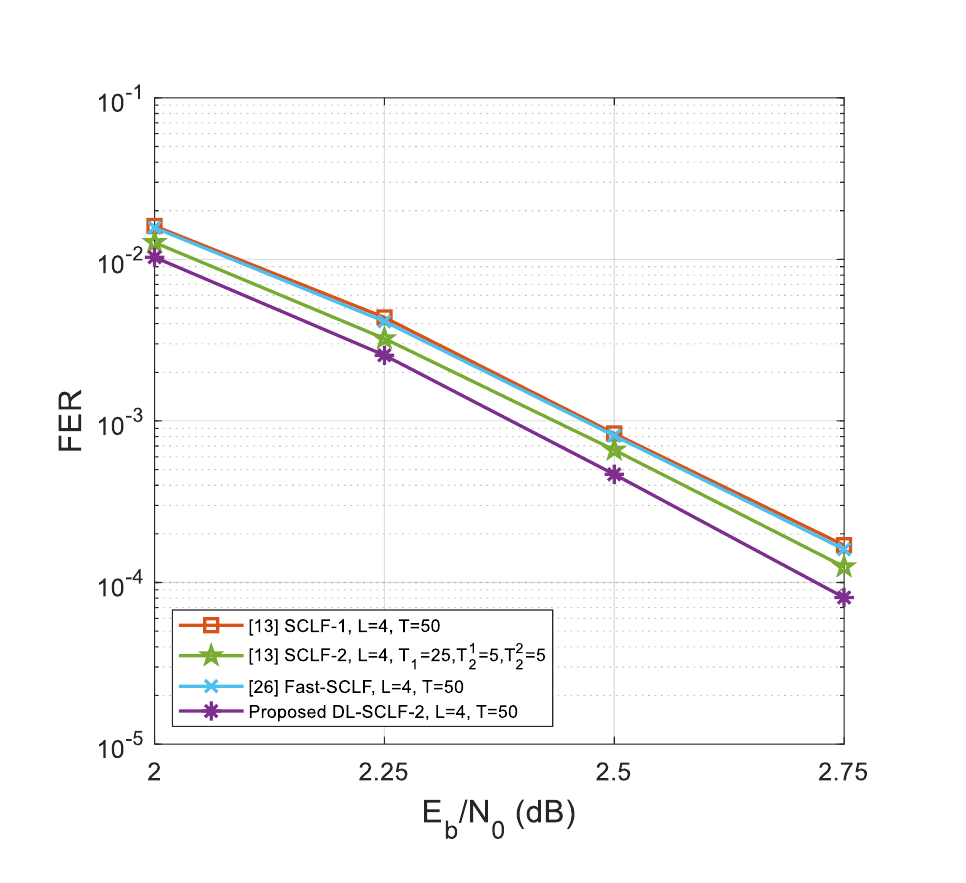}
        \label{fig:FER_proposed DL-SCLF-2 decoding_5G}
    }
    \hfill
    \subfloat[\slu{Comparison of the average number of decoding attempts}]{
        \includegraphics[page=32, width=0.45\textwidth]{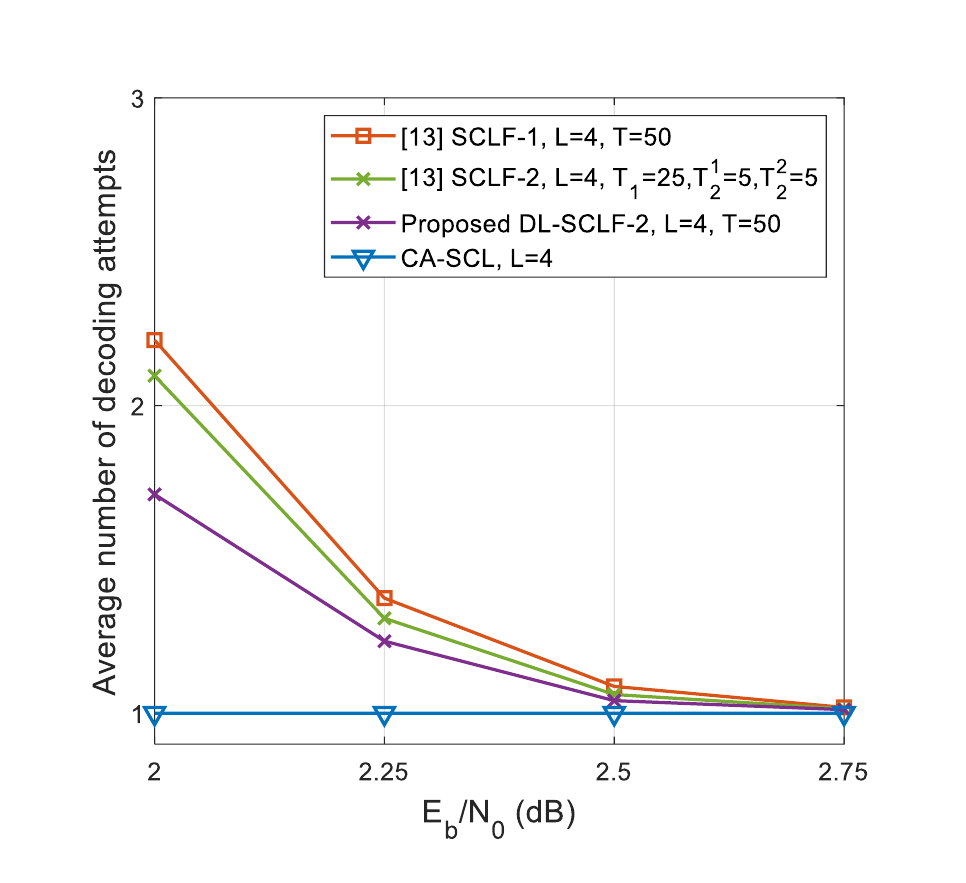}
        \label{fig:ADA_proposed DL-SCLF-2 decoding_5G}
    }
    \caption{\slu{Performance comparison for (512, 256+24) polar codes using the 5G CRC polynomial.}}
    \label{fig:proposed DL-SCLF-2 decoding_5G}
\end{figure}

\subsection{SCLF-1 Decoding Scheme}
Fig.~\ref{fig:SCLF-1 decoding} shows the FER and ADA performance of our DL-SCLF-1 decoding algorithm where the maximum number of re-decoding attempts $T=5, 10, 20$.

In terms of FER, as shown in Fig.~\ref{fig:SCLF-1 decoding}(a), we can observe that the DL-SCLF-1 decoding algorithm outperforms the other two SCLF-1 decoding algorithms proposed in \cite{CASCLF_basic_2, CASCLF_multi_bit_1}.
As the number of re-decoding attempts increases, the performance of the DL-SCLF-1 decoding algorithm is able to approach that of the genie-aided SCLF-1 decoding method. The genie-aided SCLF-1 decoding method provides the best performance that the SCLF-1 decoding technique is able to achieve, which flips the first known erroneous position in advance.

In terms of ADA, as per Fig.~\ref{fig:SCLF-1 decoding}(b), the DL-SCLF-1 decoding algorithm achieves the lowest ADA performance compared to other decoding schemes that are based on the same number of maximum re-decoding attempts.
As the SNR increases, the ADA performance will tend to match the performance achieved by CA-SCL decoding. This is because, at high SNRs, decoding is only minimally impacted by noise, so the decoding algorithms are more likely to pass the CRC detection, resulting in a lower ADA value.

\subsection{SCLF-2 Decoding Scheme}
\subsubsection{2D-DL-SCLF-2, DL-TP-SCLF-2 Decoding Comparison}
Fig.~\ref{fig:SCLF-2 decoding} compares the FER and ADA performance by the 2D-DL-SCLF-2 (\textbf{Algorithm~\ref{alg:2D-DL-SCLF-2 decoding}}) and DL-TP-SCLF-2 (\textbf{Algorithm~\ref{alg:DL-TP-SCLF-2 decoding}}) decoding schemes with the DL-SCLF-1, SCLF-1 and SCLF-2 decoding processes presented in \cite{CASCLF_multi_bit_1} based on the \textit{same total maximum number of re-decoding attempts}.
In Fig.~\ref{fig:SCLF-2 decoding}(a), the 2D-DL-SCLF-2 decoding scheme where $T_1=10$, $T^1_2=3$, and $T^2_2=3$ outperforms other SCLF decoding schemes in terms of FER performance.
This is because our stacked LSTM model improves the decoding performance by accurately identifying the position of both the first and second erroneous bits.
The FER performance for the DL-TP-SCLF-2 decoding scheme is better than that of the DL-SCLF-1 decoding scheme, but is worse than that of the 2D-DL-SCLF-2 decoding scheme based on the same number of maximum re-decoding attempts since the flipping bit position is not suitable in the DL-TP-SCLF-2 decoding algorithm.
In Fig.~\ref{fig:SCLF-2 decoding}(b), we observe that the ADA value for the 2D-DL-SCLF-2 decoding scheme is lower than that of the DL-TP-SCLF-2 decoding scheme, and both have a lower ADA value than the SCLF-1 and SCLF-2 decoding techniques proposed in \cite{CASCLF_multi_bit_1}.
Furthermore, even based on the same total maximum number of re-decoding attempts, the FER performance for the 2D-DL-SCLF-2 decoding scheme is better than that of the DL-SCLF-1 decoding approach, but more re-decoding attempts are required.

\subsubsection{Proposed DL-SCLF-2 Decoding}
Fig.~\ref{fig:proposed SCLF-2 decoding} shows the performance for the proposed DL-SCLF-2 decoding scheme (\textbf{Algorithm~\ref{alg:DL-SCLF-2 decoding:proposed method}}) compared with other decoding schemes based on the \textit{same total maximum number of re-decoding attempts}.

From the simulation illustrated in Fig.~\ref{fig:proposed SCLF-2 decoding}(a), we observe that the proposed DL-SCLF-2 decoding method outperforms the 2D-DL-SCLF-2 decoding scheme, the LSTM-SCLF-1 decoding scheme \cite{CASCLF_ml_1}, and the CA-SCL decoding scheme where $L=4$.
\slu{
Moreover, the proposed DL-SCLF-2 decoding algorithm exceeds the performance of the genie-aided SCLF-1 decoding technique and the CA-SCL decoding where $L=32$ through the post-processing of re-decoding.
As the list size of the proposed DL-SCLF-2 decoding increases, the versions with $L=16$ and $L=32$ of proposed DL-SCLF-2 decodings further approach the performance of the genie-aided SCLF-2 decoding technique.}
Fig.~\ref{fig:proposed SCLF-2 decoding}(b) illustrates that the ADA value for the proposed DL-SCLF-2 decoding method is smaller than that of 2D-DL-SCLF-2 decoding method, and is very close to that of the LSTM-SCLF-1 decoding approach \cite{CASCLF_ml_1}.
\slu{Our proposed DL-SCLF-2 decoding methods for both $L=16$ and $L=32$ demonstrate strong performance, their ADA curves are relatively flat, indicating that they are closer to one.}

Fig.~\ref{fig:diff_iter_proposed DL-SCLF-2 decoding} illustrates the performance of the proposed DL-SCLF-2 decoding method (\textbf{Algorithm~\ref{alg:DL-SCLF-2 decoding:proposed method}}) and the SCLF-2 decoding method presented in \cite{CASCLF_multi_bit_1} based on \textit{various maximum re-decoding attempts}.
As shown in Fig.~\ref{fig:diff_iter_proposed DL-SCLF-2 decoding}(a), when the maximum number of re-decoding attempts increases, the FER performance improves.
However, the performance of the SCLF-2 decoding scheme where $T_1=20$, $T^1_2=20$, and $T^2_2=20$ is almost the same as that of the SCLF-2 decoding scheme where $T_1=20$, $T^1_2=10$, and $T^2_2=10$.
This suggests that the FER performance is limited by the CRC bits and that the detection capability of the CRC-8 detector used in our simulations is insufficient.
The CRC detector may sometimes detect the wrong codeword as being the correct one before the correct codeword is decoded, causing the entire decoding process to be terminated early.
As a result, the FER performance will be saturated and cannot be improved further.
Fig.~\ref{fig:diff_iter_proposed DL-SCLF-2 decoding}(b) demonstrates that the proposed DL-SCLF-2 decoding method outperforms the SCLF-2 decoding scheme presented in \cite{CASCLF_multi_bit_1} in terms of the ADA value.

\subsection{Proposed Decoding Schemes Based on Various Code Lengths}
The following figures show simulations for the proposed DL-SCLF-2 decoding algorithm for polar codes of different code lengths.
The (256, 120+8) polar code includes a CRC-8 detector with a generator polynomial $g(x)=x^8+x^2+x+1$ and the (1024, 496+16) polar code concatenates a CRC-16 detector with a generator polynomial $g(x)=x^{16}+x^{15}+x^2+1$.

Fig.~\ref{fig:N256} shows that the proposed DL-SCLF-2 decoding scheme where $T=20$ for the (256, 120+8) polar code achieves the best performance when compared with SCLF-1 decoding algorithms \cite{CASCLF_basic_2,CASCLF_multi_bit_1} where $T=20$ and the 2D-DL-SCLF-2 decoding algorithm.
Fig.~\ref{fig:N1024} shows that the proposed DL-SCLF-2 decoding algorithm outperforms other decoding schemes in terms of both FER and ADA for the $(1024, 496+16)$ polar code where $T=50$ re-decoding attempts.

Fig.~\ref{fig:diff_iter_N1024} shows the performance comparison between the SCLF-2 decoding algorithm \cite{CASCLF_multi_bit_1} and the proposed DL-SCLF-2 decoding algorithm.
The SCLF-2 decoding algorithm \cite{CASCLF_multi_bit_1} slightly improves with an increase in $T^1_2$ and $T^2_2$.
However, the maximum number of re-decoding attempts increases exponentially.
The proposed DL-SCLF-2 decoding algorithm where $T=50$ performs in a similar manner to the SCLF-2 decoding algorithm where $T_1=50$, $T^1_2=40$, and $T^2_2=40$, and suffers a lower FER performance compared to the SCLF-2 decoding algorithm where $T_1=50$, $T^1_2=50$, and $T^2_2=50$, but has a significantly lower ADA value compared to the other two algorithms.
This indicates that the proposed DL-SCLF-2 decoding algorithm is a more effective solution as it is able to achieve better performance using fewer re-decoding attempts when compared to other decoding methods.

\subsection{\slu{Proposed DL-SCLF-2 Decoding for Polar Codes Based on 5G CRC polynomials}}
\slu{
In Fig.~\ref{fig:proposed DL-SCLF-2 decoding_5G}, we consider a (512, 256+24) polar code based on the 5G standard that is a (512, 256) polar code concatenated to a CRC detector using the generator polynomial $g(x)=x^{24}+x^{23}+x^{21}+x^{20}+x^{17}+x^{15}+x^{13}+x^{12}+x^8+x^4+x^2+x+1$.
As illustrated in Fig.~\ref{fig:proposed DL-SCLF-2 decoding_5G}, the proposed DL-SCLF-2 decoding algorithm has an advantage in terms of FER and ADA value compared to the  SCLF-1 and SCLF-2 decoding algorithms presented in \cite{CASCLF_multi_bit_1} even using the 5G CRC polynomial. Fig.~\ref{fig:proposed DL-SCLF-2 decoding_5G}(a) also includes the FER performance of the Fast-SCLF decoding algorithm presented in \cite{CASCLF_ml_5} with the same total number of re-decoding attempts. The proposed DL-SCLF-2 decoding algorithm outperforms the Fast-SCLF decoding algorithm presented in \cite{CASCLF_ml_5}. Since the proposed DL-SCLF-2 decoder is based on the conventional SCL decoding kernel while the Fast-SCLF decoder is based on the Fast-SCL decoding kernel, it is not fair to compare the decoding complexity based on the ADA. As a result, the ADA results for the Fast-SCLF decoder are not included in Fig.~\ref{fig:proposed DL-SCLF-2 decoding_5G}(b).}

\slu{For the flipping-aided SCL decoders, the ADA value gradually converges to one as the SNR increases, indicating the complexity introduced by the re-decoding attempts is negligible. In addition, we assume a built-in neural network processor existed in the future communications systems, which can be shared by different communication modules. As a result, we can just focus on  the decoding complexity of SCL decoding kernel.
The Fast-SCLF decoding \cite{CASCLF_ml_5} is designed to be hardware-friendly, significantly reducing the amount of computation. It would be interesting to explore the incorporation of the Fast-SCLF decoding algorithm into our proposed DL-SCLF-2 decoding algorithm in order to enhance both the FER performance and decoder complexity.}
\section{Conclusion}
In this paper, we consider the DL-aided SCLF decoding algorithm to improve the accuracy of the erroneous bit prediction during SCLF decoding.
First, we propose a stacked LSTM network that includes specific features to train the models to improve prediction accuracy.
Then, we separately train the models to predict the position of both the first and second erroneous bits, together with a decision about whether to continue flipping.

Based on these models, we propose DL-SCLF-1 and DL-SCLF-2 decoding algorithms. The DL-SCLF-1 decoding method is based on the basic SCLF decoding technique with the list-flipping of one-bit position, and utilizes the stacked LSTM flip-1 model to predict the position of an erroneous bit.
The proposed DL-SCLF-2 decoding scheme involves list-flipping of two bit positions based on the stacked LSTM flip-1, flip-2 model, and the continue-flipping check model.
Simulation results demonstrate that our proposed algorithms outperform existing SCLF decoding algorithms in terms of FER performance and  ADA value.

\bibliographystyle{ieeetran}

\begin{IEEEbiography}[{\includegraphics[width=1in,height=1.25in,clip,keepaspectratio]{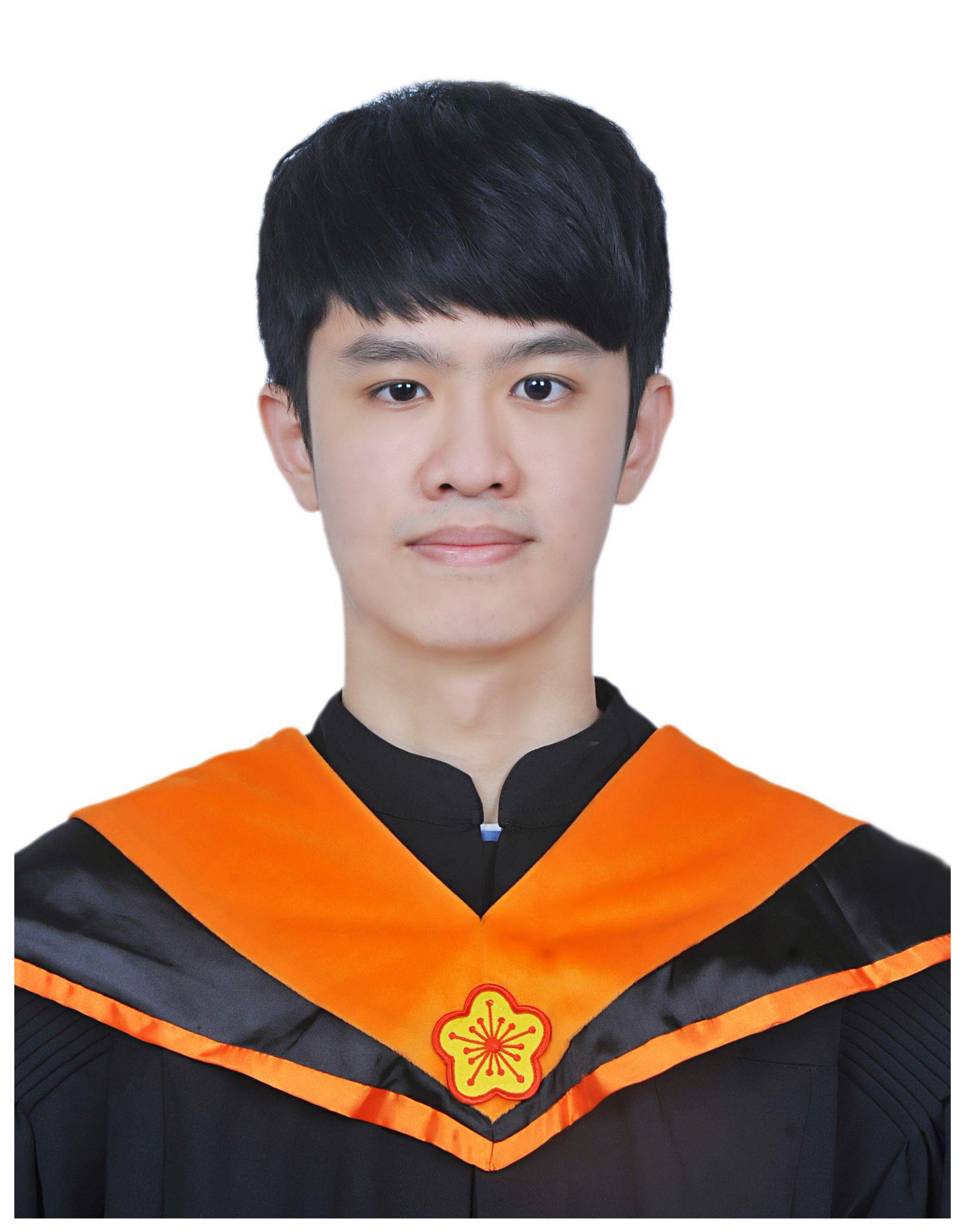}}]
	{Fu-Siang~Liang} received his B.S. degree in communication engineering from the National Chung Cheng University, Chiayi, Taiwan, in 2019, and his M.S. degree in electrical engineering from the National Tsing Hua University, Hsinchu, Taiwan, in 2023. His research interests include channel coding and deep learning.
\end{IEEEbiography}

\begin{IEEEbiography}[{\includegraphics[width=1in,height=1.25in,clip,keepaspectratio]{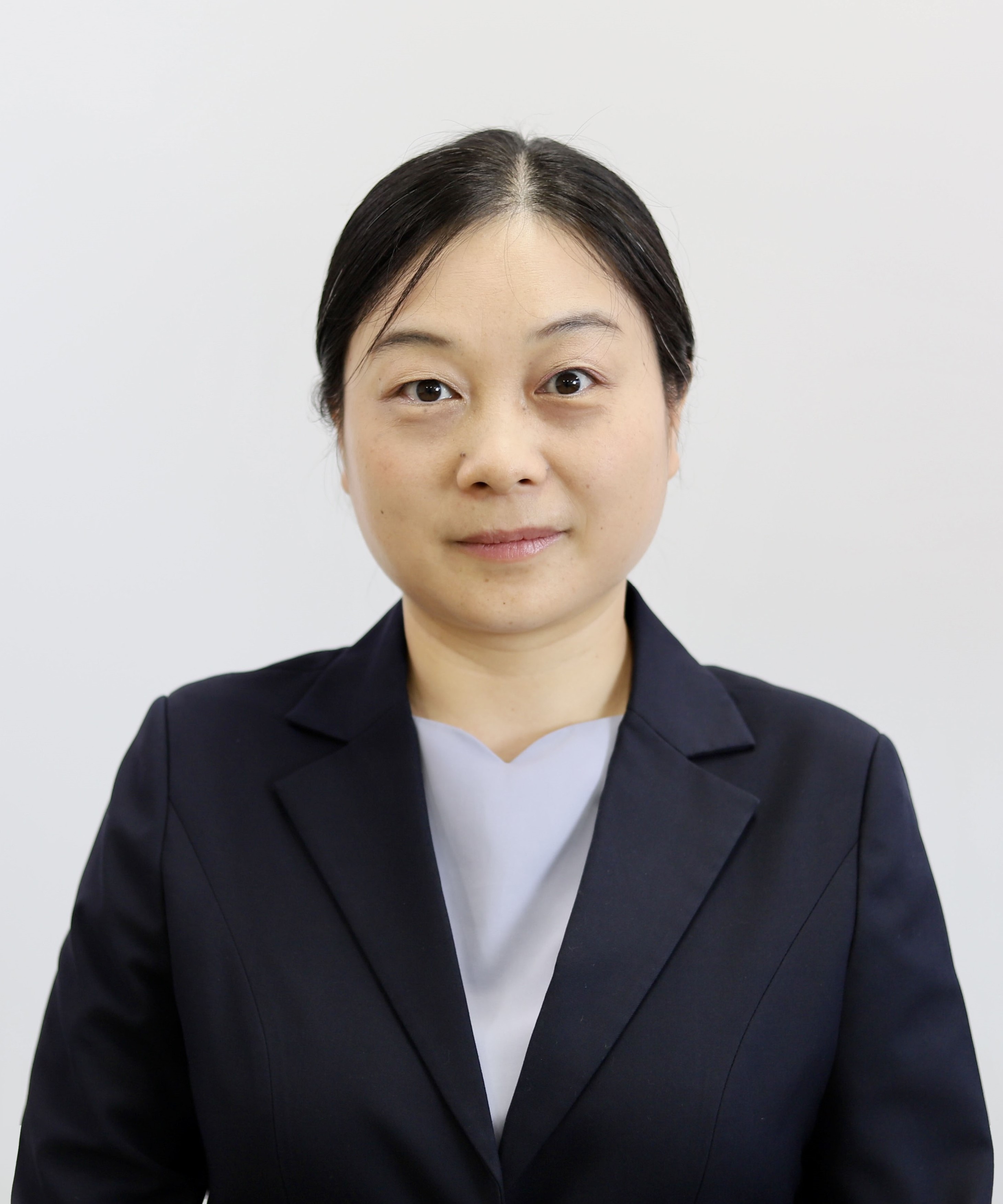}}]
	{Shan~Lu} received her B.S. and M.S. degrees in telecommunications engineering from Xidian University, Xi’an, China, in 2007 and 2010, respectively, and her Ph.D. degree in information and computer science from Doshisha University, Kyoto, Japan, in 2014. From 2014 to 2016, she was a research assistant at Doshisha University. From 2016 to 2023, she was an assistant professor at Gifu University. Currently, she is an associate professor/lecturer with the Department of Information and Communication Engineering, Graduate School of Engineering, Nagoya University, Nagoya, Japan. Her research interests are in the areas of multiuser coding, coding for nonvolatile memories, and communications theory.
\end{IEEEbiography}

\begin{IEEEbiography}[{\includegraphics[width=1in,height=1.25in,clip,keepaspectratio]{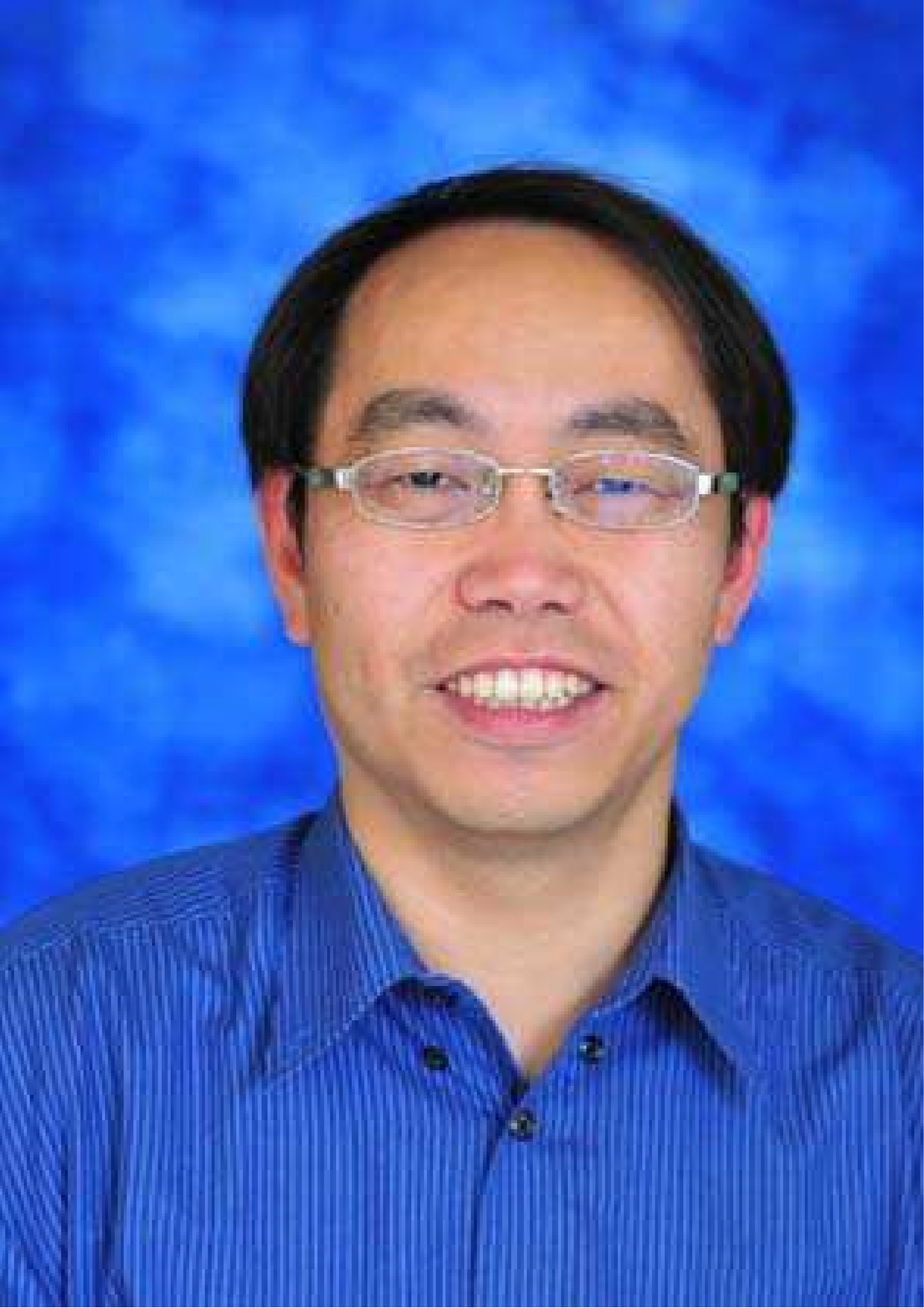}}]
	{Yeong-Luh~Ueng} (M'05-SM'15) received his Ph. D. degree in communication engineering from the National Taiwan University, Taipei, Taiwan, in 2001. From 2001 to 2005, he was with a private communication technology company, where he focused on the design and development of various wireless chips. Since December 2005, he has been a member of the faculty at the National Tsing Hua University, Hsinchu, Taiwan, where he is currently a Full Professor with the Department of Electrical Engineering and the Institute of Communications Engineering. In 2016, he was the recipient of the Wu Ta-You Memorial Award from the Ministry of Science and Technology (MOST). In 2018, he was the recipient of the Outstanding Electrical Engineering Professor award from the Chinese Electrical Engineering Association, as well as the Outstanding Research Award from the Ministry of Science and Technology (MOST), Taiwan. His research interests include coding theory, wireless communications,  communication ICs, and deep learning.
\end{IEEEbiography}

\vfill

\end{document}